\documentclass[11pt]{article}
\usepackage{axodraw}
\usepackage{epsf}
\newskip\humongous \humongous=0pt plus 1000pt minus 100pt
\def\caja{\mathsurround=0pt}
\def\eqalign#1{\,\vcenter{\openup1\jot \caja
       \ialign{\strut \hfil$\displaystyle{##}$&$
        \displaystyle{{}##}$\hfil\crcr#1\crcr}}\,}
\newif\ifdtup

\newcounter{eqnumber}[section]
\renewcommand{\theeqnumber}{\thesection.\arabic{eqnumber}}
\def\equn{
\refstepcounter{eqnumber}
\eqno({\rm \theeqnumber})
}

\def\eqn#1{eq.~{(\ref{#1})}}
\def\Eqn#1{Equation~{(\ref{#1})}}
\def\eqns#1#2{eqs.~{(\ref{#1})} and {(\ref{#2})}}
\def\fig#1{fig.~{\ref{#1}}}
\def\Fig#1{Figure~{\ref{#1}}}

\def\hf{{\textstyle{1\over2}}}
\def\eps{\epsilon}
\def\Ord{{\cal O}}
\def\I{{\cal I}}
\def\bowtie{{\rm bow\mbox{-}tie}}
\def\P{{\rm P}}
\def\NP{{\rm NP}}
\def\LC{{\rm LC}}
\def\SC{{\rm SC}}
\def\tree{{\rm tree}}
\def\oneloop{{1 \mbox{-} \rm loop}}
\def\twoloop{{2 \mbox{-} \rm loop}}
\def\mud{\lambda}
\def\Tr{\, {\rm Tr}}
\def\lr{\leftrightarrow}
\def\la{\langle}
\def\ra{\rangle}
\def\e{\epsilon}
\def\ve{\varepsilon}
\def\Li{\mathop{\rm Li}\nolimits}

\def\RS{{\scriptscriptstyle\rm R\!.S\!.}}
\def\cm{{\cal M}}
\def\bom#1{{\mbox{\boldmath $#1$}}}
\def\spa#1.#2{\left\langle#1\,#2\right\rangle}
\def\spb#1.#2{\left[#1\,#2\right]}

\def\pol{\varepsilon}
\def\si{{\sigma}}
\def\fourperm#1#2#3#4{(#1\,#2\,#3\,#4)}%

\def\cg{c_\Gamma}
\def\ctg{c_{2\Gamma}}
\def\MSbar{$\overline{\rm MS}$}

\begin{document}

\begin{titlepage}

\begin{flushright}
          SLAC--PUB--8324\\
          UCLA/99/TEP/48\\
          Saclay-SPhT-T99/147\\
          hep-ph/0001001\\
          January 1, 2000
\end{flushright}

\begin{center}
\begin{Large}
{\bf A Two-Loop Four-Gluon Helicity Amplitude in QCD} 
\end{Large}

\vskip 1.5cm

{Z. Bern$^\star$\\
\it Department of Physics and Astronomy \\
UCLA, Los Angeles, CA 90095-1547}

\vskip 0.7cm

{L. Dixon$^\dagger$\\
\it Stanford Linear Accelerator Center\\
Stanford University\\
Stanford, CA 94309}

\vskip 0.6cm
and 
\vskip 0.4cm

{D.A. Kosower\\ 
\it Service de Physique Th\'eorique$^\sharp$\\
Centre d'Etudes de Saclay\\
F-91191 Gif-sur-Yvette cedex, France} 
\end{center}

\vskip 2 cm 
\begin{abstract}
We present the two-loop pure gauge contribution to the gluon-gluon
scattering amplitude with maximal helicity violation.  Our
construction of the amplitude does not rely directly on Feynman
diagrams, but instead uses its analytic properties in $4-2\e$
dimensions.  We evaluate the loop integrals appearing in the amplitude
through $\Ord(\e^0)$ in terms of polylogarithms.
\end{abstract}


\vskip 1cm
\begin{center}
{\sl Submitted to JHEP}
\end{center}

\vfill
\noindent\hrule width 3.6in\hfil\break
\begin{small}
${}^{\star}$Research supported by the US Department of Energy under grant 
DE-FG03-91ER40662.\hfil\break
${}^{\dagger}$Research supported by the US Department of Energy under grant 
DE-AC03-76SF00515.\hfil\break
${}^{\sharp}$Laboratory of the {\it Direction des Sciences de la Mati\`ere\/}
of the {\it Commissariat \`a l'Energie Atomique\/} of France.\hfil\break
\end{small}

\end{titlepage}
             
\baselineskip 16pt


\renewcommand{\thefootnote}{\arabic{footnote}}
\setcounter{footnote}{0}


\section{Introduction}

The quest for more precise theoretical predictions for particle production
at colliders requires calculations to ever-higher orders in the
perturbative expansion of field theories. At tree level and at one loop,
workers have made substantial progress in recent years in computing
multi-parton QCD scattering amplitudes~\cite{ManganoReview,Review}.
Several important quantities, such as the total cross section for $e^+e^-$
annihilation into hadrons and the QCD $\beta$-function, have been
calculated up to four loops~\cite{FourLoopPrevious}.  In contrast,
essentially no higher-loop fixed-order results are available for
multi-parton processes depending on more than one kinematic variable.  In
particular, the two-loop amplitudes necessary for reducing the theoretical
uncertainty in the measurement of $\alpha_s$ from $e^+e^-$ jet rates and 
other event-shape variables remain uncalculated.

There has, however, been some progress in developing general formalisms 
for the computation of next-to-next-to-leading-order (NNLO) jet 
cross sections in QCD.  This includes the structure of the infrared
singularities that arise when partons are unresolved~\cite{LoopSplit},
and the explicit evaluation of two-loop integrals~\cite{Smirnov,Tausk} 
associated with massless processes.  In the special case of maximal 
$N=4$ supersymmetry, one of the authors and
his collaborators have computed the two-loop four-point amplitudes 
in terms of scalar double box integrals~\cite{BRY}.

In this paper we compute a non-supersymmetric two-loop QCD amplitude.
Our method for obtaining the amplitude is based on using
unitarity to determine its functional form.  This
program has been carried out for many one-loop and a small of number
of two-loop amplitudes~\cite{Sewing,BRY,TwoLoopGrav,GravAllN}.  The
latter computations were for the special case of theories with $N=4$
or $N=8$ supersymmetry.  The two-loop amplitude considered here is
more difficult to obtain because one cannot directly rely on
supersymmetry.  We will have to evaluate the unitarity cuts with more
care, by working in dimensional regularization with arbitrary dimension
$D=4-2\e$, taking the limit $\e \to 0$ only at the end.

The particular process we study is $g^-g^- \to g^+g^+$ in pure gauge
theory, where $g^-$ ($g^+$) denotes a negative (positive) helicity gluon.
In a helicity convention in which all gluons are treated as
outgoing, the amplitude is that for four
identical-helicity gluons, $0 \to g^+g^+g^+g^+$.  This helicity amplitude
vanishes at tree level~\cite{SWI}, and consequently the two-loop
term we compute here contributes to the cross section only at one order
beyond NNLO.  Nevertheless, the techniques used here should be applicable
to the helicity configurations that do contribute at NNLO.

This paper is organized as follows.  In section~\ref{OverviewSection} we
give an overview of the unitarity technique as applied to two-loop
amplitudes.  Section~\ref{ScalarAmplitudeSection} describes the
construction of some useful auxiliary amplitudes where scalars replace
some of the gluons circulating in the loops.  The full two-loop four-gluon
amplitude is presented in section~\ref{GluonAmplitudeSection}.  
In section~\ref{DivergenceSection} we verify that this amplitude has the
combined ultraviolet and infrared divergence structure expected on general
grounds~\cite{CataniDiv,KunsztSingular}.  An appendix contains the
expansion in $\e$, through $\Ord(\e^0)$, of the two-loop integrals 
appearing in the amplitude, expressed in terms of standard functions
(polylogarithms).

\vfill


\section{Overview of method}
\label{OverviewSection}

We reconstruct the two-loop four-gluon amplitude from its unitarity cuts, 
by evaluating the latter in dimensional regularization~\cite{HV} with
$D=4-2\e$.  Traditional applications of unitarity, via dispersion
relations, suffer in general from subtraction ambiguities.  These
ambiguities are related to the appearance of rational functions with
vanishing imaginary parts, $R(S_{i})$, where $S_{i}= \{s, t, u, \ldots\}$ 
are the kinematic variables for the amplitude.  However,
dimensionally-regulated amplitudes for massless particles necessarily
acquire a factor of $(-S_{i})^{-\e}$ for each loop, from the measure
$\int d^DL$ and dimensional analysis.  For small $\e$,
$(-S_{i})^{-\e} \, R(S_{i}) = R(S_{i}) - \e \, \ln (-S_{i}) \,
R(S_{i}) + \Ord(\e^2)$, so every term has an imaginary part (for some
$S_{i}>0$), though not necessarily in those terms which survive as
$\e\rightarrow 0$.  Thus, the unitarity cuts evaluated to $\Ord(\e)$
provide sufficient information for the complete reconstruction of an
amplitude through $\Ord(\e^0)$, subject only to the usual prescription
dependence associated with renormalization~\cite{Sewing,Review}.  The
subtraction ambiguities that arise in traditional dispersion relations
are related to the non-convergence of dispersion integrals.  A
dimensional regulator makes such integrals well-defined and
correspondingly eliminates the subtraction ambiguities. In a sense, we
use dimensional regularization as a calculational tool, beyond its
usual role as an infrared and ultraviolet regulator.

We make use of the fact that the amplitudes we are computing are also
expressible in terms of Feynman diagrams. They can therefore be
expressed as a sum of loop integrals multiplied by kinematic
coefficients.  We can therefore bypass the dispersion integrals by
identifying the loop integrals whose cuts match the appropriate products 
of tree or one-loop amplitudes.  Our main task will then be to compute the
integrands of the two-loop integrals from the tree or one-loop
amplitudes.  We obtain the full amplitude in
terms of various loop integrals, by requiring consistency of the
different cuts.  The consistency conditions are relatively simple to
implement because they hold {\it before} integrating over
intermediate-state momenta.  Because no integration is required, we
can verify the resulting form numerically (to high accuracy), even
when analytic simplification of some of the cuts is impractical.  
This approach is powerful because amplitudes on both sides of the cuts can 
be simplified, including cancellations dictated by gauge invariance, before
their products are computed.  This reduces greatly the number of terms in
intermediate steps.  The final results are quite compact when expressed in
terms of loop-momentum integrals, thus simplifying the task 
of performing the integrals explicitly.  The cut-based method thereby 
provides a short-cut to final expressions for amplitudes.
 
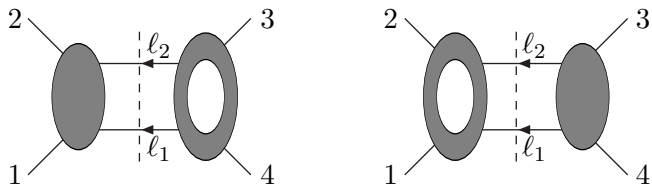
\begin{figure}[ht]
\begin{center}
\begin{picture}(250,69)(0,0)
\Text(10,5)[r]{1}  \Text(10,64)[r]{2} 
\Text(100,64)[l]{3} \Text(100,5)[l]{4}
\Text(57,15)[l]{$\ell_1$}   \Text(57,54)[l]{$\ell_2$}
\ArrowLine(82,22)(32,22) \ArrowLine(82,47)(32,47) 
\DashLine(54,10)(54,59){3.1}
\Line(29,22)(12,5) \Line(12,64)(29,47)
\Line(96,5)(79,22) \Line(79,47)(96,64)
\GOval(31,34.5)(20,10)(0){0.5}
\GOval(79,34.5)(24,12)(0){0.5}
\GOval(79,34.5)(14,7)(0){1}
%
\Text(152,5)[r]{1}  \Text(152,64)[r]{2} 
\Text(242,64)[l]{3} \Text(242,5)[l]{4}
\Text(199,15)[l]{$\ell_1$}   \Text(199,54)[l]{$\ell_2$}
\ArrowLine(224,22)(174,22) \ArrowLine(224,47)(174,47) 
\DashLine(196,10)(196,59){3.1}
\Line(171,22)(154,5) \Line(154,64)(171,47)
\Line(238,5)(221,22) \Line(221,47)(238,64)
\GOval(173,34.5)(24,12)(0){0.5}
\GOval(173,34.5)(14,7)(0){1}
\GOval(221,34.5)(20,10)(0){0.5}
\end{picture}
\caption[]{
\label{DoubleAmplFigure}
\small The $s$-channel two-particle cuts of a two-loop
amplitude as products of tree and one-loop amplitudes. We take all
external momenta to be outgoing.  The dashed lines represent the cuts.}
\end{center}
\end{figure}

Let us begin by considering the two-particle cuts of the full amplitude,
including color factors.  In the $s$ channel, in general, there are
two contributions to such cuts, as displayed in \fig{DoubleAmplFigure}.
These are,
$$
\eqalign{
&{\cal A}_4^{\twoloop}(1, 2, 3, 4) \Bigr|_{2\mbox{-}\rm cut} \cr
&\hskip0.5cm
= \sum_{\rm physical \atop states} \Bigl[
{\cal A}_4^{\tree}(1, 2, -\ell_2, -\ell_1) \times
{\cal A}_4^{\oneloop}(\ell_1, \ell_2, 3, 4)  +
{\cal A}_4^{\oneloop}(1, 2, -\ell_2, -\ell_1) \times
{\cal A}_4^{\tree}(\ell_1, \ell_2, 3, 4) \Bigr] \,, \cr}
\equn\label{TwoCut}
$$
where the notation `2-cut' means cutting the lines corresponding to
$\ell_{1,2}$, taking the absorptive part, and extracting the integrand
of the resulting phase-space integral.  On the right-hand side the
legs labeled $\ell_{1,2}$ are thus on shell in $D$ dimensions,
$\ell_i^2 =0$.  Here ${\cal A}_n^{\tree}$, 
${\cal A}_n^{\oneloop}$ and ${\cal A}_n^{\twoloop}$ are full
$n$-point amplitudes, including all color factors.  
For each external leg we have abbreviated the dependence on the
outgoing external momenta $k_i$, color indices $a_i$, and polarizations 
$\pol_i$, by the label $i$.
The sum over physical states crossing the cut contains a sum over colors.
If the cut leg is a gluon, its associated polarization vector
$\pol_{\ell_i}$ must be transverse, $\pol_{\ell_i} \cdot \ell_i = 0$.
In the sum over states, this can be imposed by a transverse projector 
$P_{\mu\nu}$ acting on the Lorentz indices crossing the cut.

\begin{figure}[ht]
\begin{center}
\begin{picture}(108,69)(0,0)
\Text(10,5)[r]{1}  \Text(10,64)[r]{2} 
\Text(100,64)[l]{3} \Text(100,5)[l]{4}
\Text(57,25)[l]{$\ell_1$}   \Text(57,41.5)[l]{$\ell_2$}
\Text(57,58)[l]{$\ell_3$}
\ArrowLine(82,18)(32,18) \ArrowLine(82,34.5)(32,34.5) 
\ArrowLine(82,51)(32,51) 
\DashLine(54,8)(54,61){3.1}
\Line(29,22)(12,5) \Line(12,64)(29,47)
\Line(96,5)(79,22) \Line(79,47)(96,64)
\GOval(31,34.5)(24,10)(0){0.5}
\GOval(79,34.5)(24,10)(0){0.5}
\end{picture}
\caption[]{
\label{TripleAmplFigure}
\small The three-particle cut of a two-loop amplitude.}
\end{center}
\end{figure}
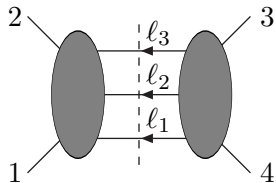

We must also consider three-particle cuts.  In the $s$-channel this
cut is displayed in \fig{TripleAmplFigure}, and is given in terms of
five-point tree amplitudes by,
$$
{\cal A}_4^{\twoloop}(1, 2, 3, 4) \Bigr|_{3\mbox{-}\rm cut} =
\sum_{\rm physical \atop states}
{\cal A}_5^{\tree}(1, 2, -\ell_3, -\ell_2, -\ell_1) \times
{\cal A}_5^{\tree}(\ell_1, \ell_2, \ell_3, 3, 4) \,,
\equn\label{ThreeCut}
$$
where the notation `3-cut' means an operation analogous to that
denoted by `2-cut', but cutting the three lines corresponding to
$\ell_{1,2,3}$.  On the right-hand side the corresponding legs are
on shell.  The two-loop amplitude will satisfy both
\eqns{TwoCut}{ThreeCut}.  These $s$-channel cuts, together with the
similar $t$- and $u$-channel ones, provide the analytic information
necessary to obtain the complete four-point two-loop amplitudes.  (The
invariants $s,t,u$ are the usual Mandelstam variables, $s=
(k_1+k_2)^2$, $t = (k_1+k_4)^2$, and $u = (k_1+k_3)^2$.)

%
\vskip .2 cm 

%
\vskip .2 cm 

The unitarity-based technique may also be viewed as an alternate way
of evaluating sets of ordinary Feynman diagrams. It does this by
collecting together gauge-invariant sets of terms which correspond to
the residues of poles in the integrands. The poles are those of the
propagators of the cut lines. This corresponds to a region of
loop-momentum integration where the cut loop momenta go on shell and
the corresponding internal lines become the intermediate states in a
unitarity relation.  From this point of view, we may consider even
more restricted regions of loop momentum integration, where additional
internal lines go on shell (and, if they are gluons, become transverse
as well).  This amounts to imposing cut conditions on additional
internal lines.

\begin{figure}[ht]
\begin{center}
\begin{picture}(156,69)(0,0)
\Text(10,5)[r]{1}  \Text(10,64)[r]{2} 
\Text(148,64)[l]{3} \Text(148,5)[l]{4}
\Text(57,15)[l]{$\ell_1$}   \Text(57,54)[l]{$\ell_2$}
\ArrowLine(82,22)(32,22) \ArrowLine(82,47)(32,47) 
\DashLine(54,10)(54,59){3.1}
\Text(105,15)[l]{$\ell_4$}   \Text(105,54)[l]{$\ell_3$}
\ArrowLine(130,22)(80,22) \ArrowLine(130,47)(80,47) 
\DashLine(102,10)(102,59){3.1}
\Line(29,22)(12,5) \Line(12,64)(29,47)
\Line(144,5)(127,22) \Line(127,47)(144,64)
\GOval(31,34.5)(20,10)(0){0.5}
\GOval(79,34.5)(20,10)(0){0.5}
\GOval(127,34.5)(20,10)(0){0.5}
\end{picture}
\caption[]{
\label{DoubleDoubleAmplFigure}
\small The $s$-channel double two-particle cut of a two-loop
amplitude separates it into a product of three tree amplitudes.  The
dashed lines represent the generalized cuts. }
\end{center}
\end{figure}
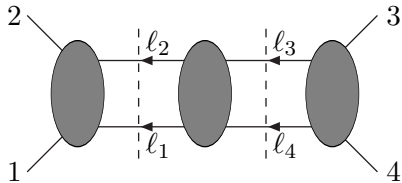

In this vein it is useful for us to define a `double' two-particle
generalized cut for a two-loop four-point amplitude.  This quantity,
illustrated in \fig{DoubleDoubleAmplFigure}, is written in terms of 
on-shell tree amplitudes as
$$
{\cal A}_4^{\twoloop} (1,2,3,4) \Bigr|_{\rm 2\times2\hbox{\small -} cut} = 
\sum_{\rm physical \atop states} {\cal A}_4^\tree(1,2, -\ell_2, -\ell_1)
\times {\cal A}_4^\tree(\ell_1, \ell_2, -\ell_3, -\ell_4) 
\times {\cal A}_4^\tree(\ell_4, \ell_3, 3, 4) \,,
\equn\label{DoubleDoubleCut}
$$
where the on-shell conditions are again imposed on the $\ell_i$,
$i=1,2,3,4$ appearing on the right-hand side.  This equation should {\it
not\/} be interpreted as trying to take `the imaginary part of an
imaginary part'. Rather it should be understood in the sense of the
previous paragraph as supplying information about the integrand of the
two-loop amplitude.  It supplies only part of the information contained in
the usual two-particle cut~(\ref{TwoCut}), which effectively imposes only
two kinematic constraints on the intermediate lines.  However, it is
simpler to evaluate because it is composed only of tree amplitudes.  Its
computation is thus a natural first step.  As we shall see this double cut
contains `most' of the final result for the amplitude we are computing in
this paper.

%
\vskip .2 cm 

Consider the two-loop four-point amplitude with all particles in the 
adjoint representation of the gauge group $SU(N_c)$.  We may write its 
color decomposition as follows~\cite{BRY},
$$
\eqalign{
{\cal A}_4^{2 \mbox{-} \rm loop}  (1,2,3,4)
 & = g^6 \hskip -.2 cm  \sum_{\si \in {S_4/Z_4}} 
N_c ^2 \Tr[T^{a_{\si(1)}}T^{a_{\si(2)}}T^{a_{\si(3)}}T^{a_{\si(4)}}]
\, \Bigl[ A^{\LC}_{4;1,1}(\si(1), \si(2), \si(3), \si(4)) \cr
& \hskip 7 cm 
     + {1\over N_c^2}
 A^{\SC}_{4;1,1}(\si(1), \si(2), \si(3), \si(4)) \Bigr] \cr
& \hskip 1 cm 
+ g^6 \hskip -.2 cm \sum_{\si \in {S_4/Z_2^3}} N_c 
\Tr[T^{a_{\si(1)}}T^{a_{\si(2)}}] \Tr[T^{a_{\si(3)}}T^{a_{\si(4)}}] \,
A_{4;1,3}(\si(1), \si(2); \si(3), \si(4))\,, \cr}
\equn\label{TwoLoopColor}
$$
where $A_{4;1,1}$ and $A_{4;1,3}$ are `partial amplitudes'.  (Our
fundamental representation color matrices are normalized by $\Tr[T^a
T^b]=\delta^{ab}$.)  The notation `$S_4/Z_4$' denotes the set of all
permutations of four objects $S_4$, omitting the cyclic transformations.
The notation `$S_4/Z_2^3$' refers again to the set of permutations of four
objects, omitting those permutations which exchange labels within a single
trace or exchange the two traces. That is, $S_4/Z_2^3 =
\{\fourperm1234,\fourperm1324,\fourperm1423\}$.  There is a similar
decomposition for contributions linearly proportional to the number of
particles in the fundamental representation, but with each term lacking one
factor of $N_c$ (see \eqn{fundscalarcolor} below).

These color decompositions have the virtue of making manifest the gauge
invariance of the partial amplitudes.  In addition the computation of
the leading-color partial amplitudes from their cuts is
straightforward.  However, the cut structure of the subleading-color
partial amplitudes is more complicated.  For this reason it will
prove convenient to use a somewhat different color decomposition,
based on structure constants $f^{abc}$ and specialized to the all-plus 
helicity amplitude, whose specification we postpone until 
section~\ref{GluonAmplitudeSection}.

In evaluating the cuts we use the 't Hooft-Veltman (HV) variant of
dimensional regularization~\cite{HV}. (For a discussion of variants of
dimensional regularization see refs.~\cite{DimRed,Long} and
refs.~\cite{KunsztFourPoint,Catani}. Note however, that the one-loop
near-equivalence of the four-dimensional helicity scheme and of
dimensional reduction assumed in refs.~\cite{KunsztFourPoint,Catani}
may not hold at two loops.)  In this scheme observed external
polarization vectors and momenta are four-dimensional, while
unobserved and internal ones are $D$-dimensional. Cut lines are
considered `internal'.  We use the convention that $D>4$ ($\e<0$) so that a
four-dimensional vector is given by setting the $(-2\eps)$-dimensional
components to zero. The HV variant violates supersymmetry Ward
identities, but is commonly used and is straightforward to implement.

In sewing the cut gluon lines we use the transverse, or physical state,
projector,
$$
P_{\mu\nu}(\ell, r) = -\eta_{\mu\nu} + 
                     {\ell_\mu r_\nu + r_\mu \ell_\nu \over r\cdot \ell}\,,
\equn\label{Projector}
$$
where $\ell$ is the gluon momentum and
$r$ is an arbitrary null `reference' momentum that drops out
of all final expressions.  The Lorentz indices in this expression are
all $D$-dimensional, but for convenience we will take $r$ to lie in
the four-dimensional subspace.  For cases where only a single gluon
crosses the cut, the second term in the projector drops out since it
is contracted with conserved currents; for cases where two or more
gluons cross the cut, the second term ensures that unphysical states do
not circulate in the loops.  We denote the number of gluon states
circulating in the loop by $D_s - 2$, which is implemented by the
Minkowski metric sum $\eta^{\mu\nu}\eta_{\mu\nu} = D_s$. 
In the HV scheme~\cite{HV} which we are using here, $D_s = D$ should be 
taken in the final expressions.  However, it is useful to keep $D_s$ 
and $D$ independent in intermediate steps.


\section{Two-loop amplitudes with an internal scalar loop}
\label{ScalarAmplitudeSection}

Although the double cut~(\ref{DoubleDoubleCut}) contains only tree
amplitudes, the gluons crossing a generalized cut carry Lorentz indices
along with them, making the algebra still non-trivial.  Contributions with
scalars crossing the cut are easier to compute first, and provide useful
information about the gluon amplitude.  Indeed at one loop, the
contributions to identical-helicity all-gluon amplitudes with a scalar
circulating in the loop are precisely equal, up to a trivial overall
constant, to the contributions with either a fermion or gluon in the
loop~\cite{OneLoopAllPlus}.  This may be understood in terms of
supersymmetry Ward identities~\cite{SWI}: ${\cal A}_n(+,+,\ldots,+)$
vanishes to any loop order in a (supersymmetrically regulated)
supersymmetric theory, and the combinations of gluon$+$fermion and
scalar$+$fermion circulating in a single loop are supersymmetric.

At two loops, even were we to use a manifestly supersymmetric regulator,
we should not expect the simple one-loop proportionality between scalar
and gluon loop contributions to persist.  Nonetheless, it is plausible
that contributions from different internal particles will be closely
related.  Indeed, for the all-plus helicity amplitude, we have found it
useful to evaluate first the contribution with an internal scalar
circulating around one of the loops. It serves as an excellent guide to an
ansatz for the pure gluon case.

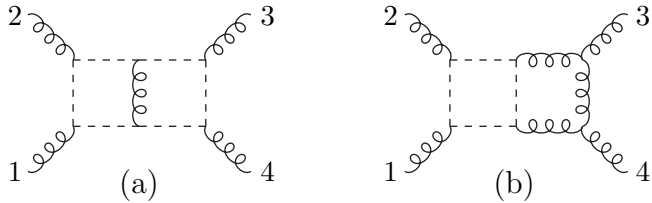
\begin{figure}[ht]
\begin{center}
\begin{picture}(250,69)(0,0)
\Text(10,5)[r]{1}  \Text(10,64)[r]{2} 
\Text(100,64)[l]{3} \Text(100,5)[l]{4}
\DashLine(79,22)(54,22){3.1} \DashLine(54,47)(79,47){3.1}
\DashLine(29,22)(54,22){3.1} \DashLine(29,47)(54,47){3.1}
\DashLine(29,22)(29,47){3.1} \Gluon(54,22)(54,47){2.5}{3}
  \DashLine(79,47)(79,22){3.1}
\Gluon(29,22)(12,5){2.5}{3} \Gluon(12,64)(29,47){2.5}{3}
\Gluon(96,5)(79,22){2.5}{3} \Gluon(79,47)(96,64){2.5}{3}
\Text(54,0)[c]{\large (a)}
\Text(152,5)[r]{1}  \Text(152,64)[r]{2} 
\Text(242,64)[l]{3} \Text(242,5)[l]{4}
\Gluon(221,22)(196,22){2.5}{3} \Gluon(196,47)(221,47){2.5}{3}
\DashLine(171,22)(171,47){3.1} \DashLine(171,22)(196,22){3.1}
\DashLine(171,47)(196,47){3.1}
\DashLine(196,22)(196,47){3.1}  \Gluon(221,47)(221,22){2.5}{3}
\Gluon(171,22)(154,5){2.5}{3} \Gluon(154,64)(171,47){2.5}{3}
\Gluon(238,5)(221,22){2.5}{3} \Gluon(221,47)(238,64){2.5}{3}
\Text(196,0)[c]{\large (b)}
\end{picture}
\caption[]{
\label{ParentScalarLoopFigure}
\small Representative diagrams for the contributions with a 
single scalar loop.  The class (a) diagrams can be drawn on
the plane with all external legs on the outside, and one internal 
gluon line on the inside of the scalar loop.  The class (b) diagrams
have all the gluon lines on the outside of the scalar loop. }
\end{center}
\end{figure}

To illustrate the utility of scalar-loop contributions in constructing the
gluon-loop one, consider the two-loop four-gluon amplitudes with a single
scalar loop.  \Fig{ParentScalarLoopFigure} shows that there are two
different classes of planar diagrams, divided according to whether or not
an internal gluon line runs inside the scalar loop.
For the case of scalars in the fundamental representation of $SU(N_c)$,
the contributions proportional to the color trace 
$\Tr[T^{a_1} T^{a_2} T^{a_3} T^{a_4}]$ are
$$
\Tr[T^{a_1} T^{a_2} T^{a_3} T^{a_4}] \,
\Bigl[ N_c A_4^{\twoloop, \rm (b)} 
- {1\over N_c} A_4^{\twoloop, \rm (a)} \Bigr] \,,
\equn\label{fundscalarcolor}
$$
where $N_c$ is the number of colors, and the labels (a) and (b) correspond
to the diagram classes depicted in \fig{ParentScalarLoopFigure}.  The
amplitudes are color-stripped; that is, they should be computed using
color-ordered Feynman rules~\cite{ManganoReview} from which color factors
have been removed.  Because the diagrams of class (a) and (b) enter
\eqn{fundscalarcolor} with different powers of $N_c$, they are separately
gauge invariant and may be computed independently.  For scalars in the
adjoint representation the two contributions are still separately gauge
invariant, but there is no relative factor of $-N_c^2$. The leading-color
contribution proportional to the same color trace is then
$$
\Tr[T^{a_1} T^{a_2} T^{a_3} T^{a_4}] \,
N_c^2 \Bigl[A_4^{\twoloop, \rm (b)} + A_4^{\twoloop, \rm (a)} \Bigr] \,.
\equn\label{ScalarAdjoint}
$$

We begin our evaluation of scalar-loop contributions with the $s$-channel 
double two-particle cut for the subleading-color term in
\fig{ParentScalarLoopFigure}(a).  This is given by the product
(\ref{DoubleDoubleCut}) where only scalars cross the
generalized cuts,
$$
\eqalign{
A_4^{\twoloop, \rm (a)} &(1^+,2^+,3^+,4^+) 
\Bigr|_{\rm 2\times2\hbox{\small -} cut}  \cr
& = 
 A_4^\tree(1^+, 2^+, -\ell_2^s, -\ell_1^s) \times  
 A_4^\tree(\ell_1^s, \ell_2^s, -\ell_3^s, -\ell_4^s) \times
 A_4^\tree(\ell_4^s, \ell_3^s, 3^+, 4^+) \cr
& = i {\spb1.2 \spb3.4 \over \spa1.2 \spa3.4}
\biggl( {s_{12} \mud_1^2 \mud_4^2 \over
   (\ell_1 - k_1)^2 (\ell_1 - \ell_4)^2 (\ell_4+k_4)^2} 
+ {1\over 2} {\mud_1^2 \mud_4^2 \over (\ell_1 - k_1)^2 (\ell_4+k_4)^2 }
 \biggr) \,, \cr}
\equn\label{DoubleDoubleScalar}
$$
where $s_{ij} = (k_i + k_j)^2$. The vectors $\vec\mud_i$ represent the
$(-2\eps)$-dimensional components of the loop momenta $\ell_i$; 
that is, $\ell_i \equiv \ell_i^{[4]} + \mud_i$, where
$\ell_i^{[4]}$ has only four-dimensional components.
Superscripts $\pm$ label helicities of external gluons, while a 
superscript $s$ indicates that a cut leg is a scalar.  (A cut 
momentum without a superscript will implicitly denote a gluon.)
We used the color-ordered~\cite{ManganoReview} tree amplitudes
(or partial amplitudes),
$$
\eqalign{
& A_4^\tree(1^+, 2^+, -\ell_2^s, -\ell_1^s) = 
i \mud_1^2 {\spb1.2 \over \spa1.2} 
 {1 \over (\ell_1 - k_1)^2}\,, \cr
& A_4^\tree(\ell_1^s, \ell_2^s, -\ell_3^s, -\ell_4^s) =
 -i \biggl({s_{12} \over (\ell_1 - \ell_4)^2} + {1\over 2} \biggr) \,; \cr}
\equn\label{DoubleDoubleTrees}
$$
$A_4^\tree(\ell_4^s,\ell_3^s, 3^+, 4^+)$ can be obtained
by relabeling  $A_4^\tree(1^+, 2^+, -\ell_2^s, -\ell_1^s)$.
We have expressed the amplitudes in terms of spinor inner
products~\cite{ManganoReview} which are denoted by $\spa{i}.j = \la
i^- | j^+\ra$ and $\spb{i}.j = \la i^+| j^-\ra$, where $|i^{\pm}\ra$
are four-dimensional massless Weyl spinors of momentum $k_i$, labeled 
with the sign of the helicity. 

\begin{figure}[ht]
\begin{center}
\begin{picture}(170,69)(0,0)
\Text(3,10)[r]{1} \Text(3,50)[r]{2} 
\Line(5,10)(13,18)  \Line(5,50)(13,42)  
\Line(57,18)(65,10)  \Line(57,42)(65,50) 
\Line(13,18)(57,18)  \Line(13,42)(57,42) 
\Line(13,18)(13,42) \Line(35,18)(35,42) \Line(57,18)(57,42) 
\Text(67,50)[l]{3} \Text(67,10)[l]{4} 
\Text(35,0)[c]{\large (a)}
\Text(103,10)[r]{1} \Text(103,50)[r]{2} 
\Line(105,10)(137,42)  \Line(105,50)(124,31) \Line(126,29)(137,18)  
\Line(157,18)(165,10)  \Line(157,42)(165,50) 
\Line(113,18)(157,18)  \Line(113,42)(157,42) \Line(157,18)(157,42) 
\Text(167,50)[l]{3} \Text(167,10)[l]{4} 
\Text(135,0)[c]{\large (b)}
\end{picture}
\caption[]{
\label{ParentsFigure}
\small The planar and non-planar double box integrals.}
\end{center}
\end{figure}
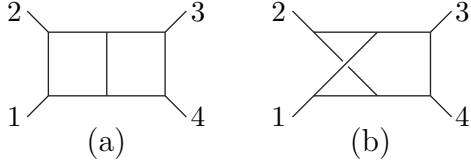

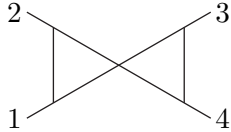
\begin{figure}[ht]
\begin{center}
\begin{picture}(79.3,69)(0,0)
\Text(3,10)[r]{1} \Text(3,50)[r]{2} 
\Line(5,10)(74.3,50)  \Line(5,50)(74.3,10)  
\Line(15,15.77)(15,44.23) \Line(64.3,15.77)(64.3,44.23)
\Text(76.3,50)[l]{3} \Text(76.3,10)[l]{4} 
\end{picture}
\caption[]{
\label{BowTieFigure}
\small The `bow-tie' integral appearing in \eqn{ScalarAnsatzA}.}
\end{center}
\end{figure}

As a working hypothesis, from \eqn{DoubleDoubleScalar} we take
the subleading-color planar amplitude to be%
\footnote{These bare amplitudes should be multiplied by the scale factor
$(\mu_0^2)^{2\eps}$, and after renormalization instead by the scale 
$(\mu_R^2)^{2\eps}$. Here we take these scales to equal unity except 
where noted.} 
$$
\eqalign{
A_4^{\twoloop,\rm (a)} (1^+,2^+,3^+,4^+) & = 
i {\spb1.2 \spb3.4 \over \spa1.2 \spa3.4} \Bigl[
 s_{12} \I_4^\P[\mud_p^2 \mud_q^2](s_{12}, s_{23})  
      + {1\over 2} \I_4^\bowtie[\mud_p^2 \mud_q^2](s_{12}) \cr
& \hskip 5 cm   + (s_{12} \leftrightarrow s_{23}) \Bigr] \,,\cr}
\equn\label{ScalarAnsatzA}
$$
where the two terms correspond to the two terms in the second equation
of \eqn{DoubleDoubleScalar},
$$
\eqalign{
\I_4^\P [{\cal P} & (\mud_i, p,q,k_i)] (s_{12},s_{23}) \cr
 & \equiv \int
 {d^{D}p\over (2\pi)^{D}} \,
 {d^{D}q\over (2\pi)^{D}}\,
 { {\cal P} (\mud_i, p,q,k_i) \over 
     p^2\, q^2\, (p+q)^2 (p - k_1)^2 \,(p - k_1 - k_2)^2 \,
        (q - k_4)^2 \, (q - k_3 - k_4)^2 }  \cr}
\equn\label{PlanarInt} 
$$
is the planar double box integral displayed in \fig{ParentsFigure}(a),
and the vectors $\vec\mud_p$, $\vec\mud_q$ represent
the $(-2\eps)$-dimensional components of the loop momenta $p$ and $q$.
The numerator factor ${\cal P} (\mud_i, p,q,k_i)$ is a
polynomial in the momenta.
The `bow-tie' integral $\I_4^\bowtie$ shown in \fig{BowTieFigure} is
defined by
$$
\eqalign{
\I_4^{\bowtie} [{\cal P} & (\mud_i, p,q,k_i)] (s_{12}) \cr
 & \equiv \int
 {d^{D}p\over (2\pi)^{D}} \,
 {d^{D}q\over (2\pi)^{D}}\,
 { {\cal P} (\mud_i, p,q,k_i) \over 
     p^2\, q^2\, (p - k_1)^2 \,(p - k_1 - k_2)^2 \,
        (q - k_4)^2 \, (q - k_3 - k_4)^2 }\,. \cr}
\equn\label{BowTieInt} 
$$
The one `missing' propagator, compared with $\I_4^\P$, makes
$\I_4^{\bowtie}$ much simpler to evaluate, since it is a product of two
independent one-loop integrals.  Although it is not manifest, the overall 
spinor prefactor is symmetric under all 4! permutations of the external 
legs, including the cyclic relabeling, 
$k_1 \rightarrow k_2 \rightarrow k_3 \rightarrow k_4 \rightarrow k_1$. 
Hence the amplitude~(\ref{ScalarAnsatzA}), including the term 
with $s_{12}$ and $s_{23}$ exchanged, is cyclicly invariant too.
The exchange term generates the proper double two-particle cut in the 
$t$-channel.

%

%

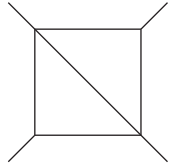
\begin{figure}[ht]
\begin{center}
\begin{picture}(70,70)(0,0)
\Line(5,65)(65,5) \Line(5,5)(15,15) \Line(55,55)(65,65)
\Line(15,15)(55,15) \Line(55,15)(55,55) 
\Line(55,55)(15,55) \Line(15,55)(15,15)
\end{picture}
\caption[]{
\label{ThreePartFuncFigure}
\small An integral function containing only three particle cuts.}
\end{center}
\end{figure}

Cutting a line forces the corresponding propagator to be present,
uncanceled, in all integral functions.  In general, an ansatz for an
amplitude based on a limited set of cuts will fail to capture integral
functions which are missing some or all of the cut propagators. As one
refines the ansatz by investigating additional cuts, it may be
necessary to add in integral functions that do not have cuts in
previously-investigated channels.  For example, the integral depicted
in \fig{ThreePartFuncFigure} contains only three-particle cuts and
would not be detected in two-particle cuts.  However, for the planar
amplitude under consideration, it turns out that such integrals do not
appear, and the double two-particle cuts are in fact sensitive to all
functions appearing in the amplitude.

%

\begin{figure}[ht]
\begin{center}
\begin{picture}(207,69)(0,0)
\Text(15,5)[r]{1\ } \Text(15,64)[r]{2\ }
\Text(73,64)[l]{\ 3} \Text(73,5)[l]{\ 4}
\Text(45,-2)[c]{\large (a)}
\DashLine(56,22)(31,22){3.1} \DashLine(31,47)(56,47){3.1}
\DashLine(31,22)(31,47){3.1} \Gluon(56,47)(56,22){2.5}{3}
\Gluon(31,22)(14,5){2.5}{3} \Gluon(14,64)(31,47){2.5}{3}
\DashLine(56,22)(73,5){3.1} \DashLine(56,47)(73,64){3.1}
\Text(135,5)[r]{1\ } \Text(135,64)[r]{2\ }
\Text(193,64)[l]{\ 3} \Text(193,5)[l]{\ 4}
\Text(165,-2)[c]{\large (b)}
\Gluon(176,22)(151,22){2.5}{3} \Gluon(151,47)(176,47){2.5}{3}
\Gluon(151,22)(151,47){2.5}{3} \DashLine(176,47)(176,22){3.1}
\Gluon(151,22)(134,5){2.5}{3} \Gluon(134,64)(151,47){2.5}{3}
\DashLine(176,22)(193,5){3.1} \DashLine(176,47)(193,64){3.1}
\end{picture}
\caption[]{
\label{OneLSublScalarFigure}
\small Representative planar diagrams for the 
subleading-in-color (a) and leading-in-color (b) one-loop amplitudes 
for two gluons and two scalars.  The diagrams are divided into the two
classes according to whether or not there is an internal gluon line to 
the right of the scalar line. }
\end{center}
\end{figure}
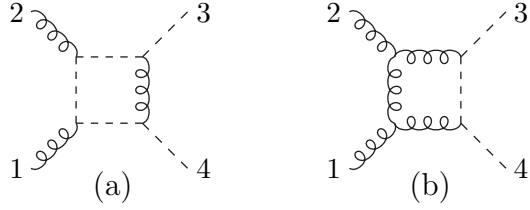

To verify that \eqn{ScalarAnsatzA} is indeed the complete amplitude
corresponding to the class (a) diagrams in \fig{ParentScalarLoopFigure} 
we must evaluate the two- and three-particle cuts.  To evaluate the 
two-particle cuts we need the first of the tree amplitudes given
in \eqn{DoubleDoubleTrees}, as well as the subleading-in-color 
one-loop amplitude (see \fig{OneLSublScalarFigure}(a) for 
a representative diagram),
$$
 A_4^{\oneloop, \rm (a)}(1^+,2^+,3^s,4^s)
 = - {\spb1.2\over\spa1.2} \left[ 
    s_{12} \, \I_4^\oneloop[\mud_p^2] (s_{12},s_{23})
   + {1\over2} \, \I_3^{\oneloop,(4)}[\mud_p^2](s_{12}) \right] \,.
\equn\label{ppssOneLoop}
$$
(We compute all required one-loop amplitudes from their
$s$- and $t$-channel cuts.)  We have taken the momenta of the external
legs 1 and 2 to be four-dimensional while the momenta of legs 3 and 4
are $D$-dimensional, so that they can be inserted into the two-particle
cuts.  The one-loop box integral in \eqn{ppssOneLoop} is defined by
$$
\I_4^\oneloop[ {\cal P} (\mud_p,p,k_i)] \equiv 
 \int {d^{D}p\over (2\pi)^{D}} \,  { {\cal P} (\mud_p, p,k_i) \over 
       p^2\,(p - k_1)^2 \,(p - k_1 - k_2)^2 \, (p + k_4)^2}\,.
\equn\label{BoxIntegral}
$$
The triangle integral $\I_3^{\oneloop,(4)}$ is obtained
by removing the $1/(p+k_4)^2$ propagator from \eqn{BoxIntegral}.  
Taking the two-particle cut of the ansatz (\ref{ScalarAnsatzA}) we obtain
$$
\eqalign{
A_4^{\twoloop, (a)} (1^+,2^+,3^+,4^+) 
\Bigr|_{\rm 2\hbox{\small -} cut} & = 
 A_4^\tree(1^+, 2^+, -\ell_2^s, -\ell_1^s) \times  
 A_4^{\oneloop, \rm (a)}(\ell_2^s, \ell_1^s, 3^+, 4^+) \cr
& +  A_4^{\oneloop, \rm (a)}(1^+, 2^+, -\ell_2^s, -\ell_1^s) \times  
 A_4^\tree(\ell_2^s, \ell_1^s, 3^+, 4^+) \,, \cr} 
\equn
$$
as required.  The two-particle cuts work simply in this example because
all terms in the one-loop amplitude~(\ref{ppssOneLoop}) could be detected
via the $s$-channel cut.

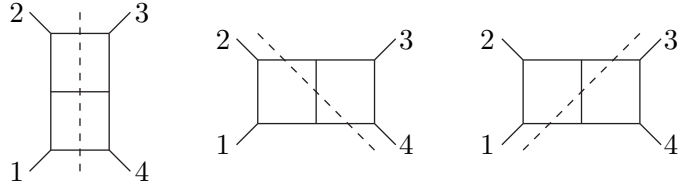
\begin{figure}[ht]
\begin{center}
\begin{picture}(300,100)(0,0)
\Text(25,0)[r]{1} \Text(25,60)[r]{2} 
\Line(35,30)(57,30) \Line(35,8)(57,8) \Line(35,52)(57,52)
\Line(27,0)(35,8) \Line(65,0)(57,8)
\Line(27,60)(35,52) \Line(65,60)(57,52)
\Line(35,8)(35,52) \Line(57,8)(57,52)
\DashLine(46,0)(46,60){3}
\Text(67,60)[l]{3} \Text(67,0)[l]{4} 
\Text(103,10)[r]{1} \Text(103,50)[r]{2} 
\Line(105,10)(113,18)  \Line(105,50)(113,42)  
\Line(157,18)(165,10)  \Line(157,42)(165,50) 
\Line(113,18)(157,18)  \Line(113,42)(157,42) 
\Line(113,18)(113,42) \Line(135,18)(135,42) \Line(157,18)(157,42) 
\DashLine(113,52)(157,8){3}
\Text(167,50)[l]{3} \Text(167,10)[l]{4} 
\Text(203,10)[r]{1} \Text(203,50)[r]{2} 
\Line(205,10)(213,18)  \Line(205,50)(213,42)  
\Line(257,18)(265,10)  \Line(257,42)(265,50) 
\Line(213,18)(257,18)  \Line(213,42)(257,42) 
\Line(213,18)(213,42) \Line(235,18)(235,42) \Line(257,18)(257,42) 
\DashLine(213,8)(257,52){3}
\Text(267,50)[l]{3} \Text(267,10)[l]{4} 
\end{picture}
\caption[]{
\label{TripleCutFigure}
\small The $s$-channel three-particle cuts of the class (a) scalar
contribution in \eqn{ScalarAnsatzA}, and of the leading-color 
pure gluon contribution in \eqn{LeadingGlueResult}.  
The dashed lines represent the cuts.}
\end{center}
\end{figure}

The three-particle cuts are more intricate, but may be handled
numerically.  For the $s$-channel case, we need the product of tree
amplitudes,
$$
A_5^\tree(1^+, 2^+, -\ell_3^s, -\ell_2^\mu, -\ell_1^s) \times 
P_{\mu\nu}(\ell_2, r)
\times  A_5^\tree(\ell_1^s, \ell_2^\nu, \ell_3^s, 3^+, 4^+) \,,
\equn\label{TwoL3ParticleCut}
$$
where $\mu,\nu$ are Lorentz indices for the single gluon line crossing 
the cut, and the physical state projector $P_{\mu\nu}$ is defined in
\eqn{Projector}.  If our ansatz~(\ref{ScalarAnsatzA}) is correct, then
the cut expression~(\ref{TwoL3ParticleCut}) should be equal to
$$
\eqalign{
 & i {\spb1.2\spb3.4 \over \spa1.2 \spa3.4} 
    \mud_1^2\mud_3^2 
 \biggl[  {s_{23} \over (\ell_1-k_1)^2 (\ell_3-k_2)^2 (\ell_3+k_3)^2 
             (\ell_1+k_4)^2} \cr
& \hskip 3.2 cm 
        +{s_{12} \over (\ell_1+\ell_2)^2 (\ell_3-k_2)^2 (\ell_2+\ell_3)^2 
             (\ell_1+k_4)^2} 
          \cr
& \hskip 3.2 cm 
        +{s_{12} \over (\ell_1+\ell_2)^2 (\ell_3+k_3)^2 (\ell_2+\ell_3)^2 
       (\ell_1-k_1)^2} \biggr] \,, }
\equn\label{AThreeCut}
$$
which is obtained by cutting the two planar double box integrals in
\eqn{ScalarAnsatzA} as shown in \fig{TripleCutFigure}.  The bow-tie
integrals do not contribute to three-particle cuts.  The equality of these
expressions should hold before any loop or phase-space integration is
performed.

%

The expressions for the cuts are smooth analytic functions with no
explicit dependence on the dimension $D$ of the loop-momentum vectors.  It
is therefore sufficient to verify the cuts for integer values of $D$, with
$D>4$, where numerical evaluation is straightforward.  We randomly
generate $D$-vectors satisfying the appropriate cut kinematics --- in this
case, seven null-vectors $\{k_1,k_2,k_3,k_4,\ell_1,\ell_2,\ell_3\}$ obeying
$$
k_1+k_2-\ell_1-\ell_2-\ell_3\ =\ k_3+k_4+\ell_1+\ell_2+\ell_3\ =\ 0,
\equn
$$ 
where the $k_i$ are four-dimensional, i.e., they have vanishing
extra-dimensional components.  We then check whether the cuts agree with
our ansatz at this point in phase space, in this case by comparing
numerical values for \eqns{TwoL3ParticleCut}{AThreeCut}.  Because no
integration is required, the check can be performed to very high accuracy
(25 or more digits), and repeated quickly for other phase-space points.
As mentioned previously, we take the dimension $D_s$, which appears in the
number of gluon states circulating in the loop, to be independent from the
dimension $D$ of the loop momentum vectors.  To distinguish between
e.g. $(\mud_1\cdot \mud_2)^2$ and $\mud_1^2 \mud_2^2$ (the most
subtle degeneracy arising at two loops) it is necessary to choose 
$D \ge 6$.

Together with the two-particle cuts, the numerical verification of the 
three-particle cuts proves that the ansatz~(\ref{ScalarAnsatzA})
does indeed give the complete subleading-color contribution to the 
color coefficient $\Tr[T^{a_1} T^{a_2} T^{a_3} T^{a_4}]$ for the
case of a scalar in the loop.


Following a similar procedure we have also obtained the results 
for the leading-color contributions, obtained from the diagrams 
of the type displayed in \fig{ParentScalarLoopFigure}(b).  
The color-stripped amplitude for case (b) is, 
$$
\eqalign{
A_4^{\rm 2-loop, \rm (b)}(1^+,2^+,3^+,4^+) &
= i \, { \spb1.2\spb3.4 \over \spa1.2\spa3.4 } \biggl[
  s_{12} \, \I_4^\P[ \mud_p^2 \, \mud_{p+q}^2 ](s_{12},s_{23})  \cr 
& \hskip .5 cm
+ {(D_s-2) \over s_{12}} \, \I_4^\bowtie[ 
      \mud_p^2 \, \mud_q^2 \, ( (p+q)^2 + s_{12} ) ] (s_{12},s_{23}) \cr
& \hskip .5 cm
+ 4 \, \I_4^\bowtie[ \mud_p^2 \, (\mud_p \cdot \mud_q) ] (s_{12}) 
+ \hbox{cyclic perms of }(1,2,3,4)  \biggr] \,. \cr}
\equn\label{ScalarAnsatzB}
$$
In this case the number of gluonic states $(D_s-2)$ circulating in the
loop appears.  Although some of the terms in \eqn{ScalarAnsatzB}
vanish under integration, we have kept them because they do not
vanish in the cuts before integration and are therefore important for
verifying the cuts numerically.

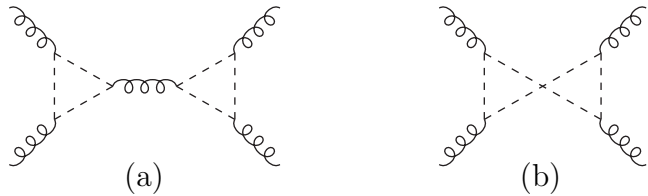
\begin{figure}[ht]
\begin{center}
\begin{picture}(230,69)(0,0)
\Text(46,0)[c]{\large (a)}
\DashLine(12,22)(12,47){3.1} \DashLine(12,47)(34,34.5){3.1}
  \DashLine(34,34.5)(12,22){3.1}
\DashLine(80,47)(80,22){3.1} \DashLine(80,22)(58,34.5){3.1}
  \DashLine(58,34.5)(80,47){3.1}
\Gluon(34,34.5)(58,34.5){2.5}{3}
\Gluon(12,22)(-5,5){2.5}{3} \Gluon(-5,64)(12,47){2.5}{3}
\Gluon(97,5)(80,22){2.5}{3} \Gluon(80,47)(97,64){2.5}{3}
\Text(196,0)[c]{\large (b)}
\DashLine(174,22)(174,47){3.1} \DashLine(218,47)(218,22){3.1} 
\DashLine(218,22)(174,47){3.1} \DashLine(174,22)(218,47){3.1}
\Gluon(174,22)(157,5){2.5}{3} \Gluon(157,64)(174,47){2.5}{3}
\Gluon(235,5)(218,22){2.5}{3} \Gluon(218,47)(235,64){2.5}{3}
\end{picture}
\caption[]{
\label{DoubleScalarFigure}
\small Representative planar diagrams with two scalar loops.  
The ones in (a) have two distinct loops connected by a gluon line;
the ones in (b) have a quartic scalar interaction. }
\end{center}
\end{figure}

The two-loop four-gluon amplitude with scalars present can also receive 
contributions with two scalar loops.  \Fig{DoubleScalarFigure}(a) shows
a class that is present just due to the gauge interaction.
Because supersymmetric gauge theories with scalars always include quartic
scalar interactions (arising from $D$-terms), it is useful to also
calculate the class shown in \fig{DoubleScalarFigure}(b).
Both types of contributions are simple,
$$
\eqalign{
A_4^{\rm double-scalar,(a)}(1^+,2^+,3^+,4^+)\  
&\propto\ 
i \, { \spb1.2\spb3.4 \over \spa1.2\spa3.4 } \, {1 \over s_{12}} \,
   \I_4^\bowtie[ \mud_p^2 \, \mud_{q}^2 \, 
    ( (p+q)^2 + \hf s_{12} ) ](s_{12},s_{23}) \,,  \cr 
A_4^{\rm double-scalar,(b)}(1^+,2^+,3^+,4^+)\  
&\propto\  
i \, { \spb1.2\spb3.4 \over \spa1.2\spa3.4 } \,
   \I_4^\bowtie[ \mud_p^2 \, \mud_{q}^2 ](s_{12}) \,. \cr }
\equn\label{ScalarAnsatzC}
$$
The precise normalization is unimportant; in constructing the pure gluon
ansatz we shall add these terms to \eqns{ScalarAnsatzA}{ScalarAnsatzB}
with coefficients to be determined later.

\section{The pure gluon amplitude}
\label{GluonAmplitudeSection}

We now turn to the two-loop pure gluon amplitude 
${\cal A}_4^\twoloop(1^+,2^+,3^+,4^+)$, which is the main subject of this
paper.  It is convenient to begin with the leading-color contribution
generated by planar graphs.

\subsection{Leading color}
\label{LeadingGluonSection}

The planar amplitudes with internal scalars from 
section~\ref{ScalarAmplitudeSection} serve as a guide toward constructing
the planar pure gluon contributions.  We write an ansatz in terms of
loop integrals by taking a linear combination of the above results for one 
scalar in the loop, plus contributions with two independent 
scalar loops.  This turns out not to be the complete answer so an additional 
polynomial in the $\mud_i$ is required, whose form is not too hard to guess.
We numerically evaluate the $D$-dimensional cuts at a number of different
phase-space points, and use this information to solve for the unknown 
(but kinematically constant) coefficients of possible additional terms.  
In this way we obtain a simple representation of the planar pure gluon 
amplitude in terms of loop integrals.  The form of the planar amplitude 
then suggests an ansatz for remaining non-planar contributions, which we 
again verify numerically from the cuts.

In somewhat more detail, our starting point is the contribution of a
single adjoint scalar loop, \eqn{ScalarAdjoint}.  We multiply this
contribution by the number of gluon states, $D_s-2$. This would produce
the complete answer at one loop, but here we must make two corrections:
\begin{itemize}
\item a factor of $1/2$ for terms proportional to $(D_s-2)^2$, from 
different combinatorics of contributions in which the two loops have 
no common propagator, as in \fig{DoubleScalarFigure}(a).
\item the addition of a term (with an unknown coefficient) induced by 
a quartic scalar interaction, as in \fig{DoubleScalarFigure}(b).  
\end{itemize}
Even with these corrections, the ansatz does not quite work.  However, it
does satisfy the double two-particle and three-particle cuts in $D=4$
(i.e., when all $(-2\e)$-dimensional components are set to zero) and also
in $D=5$.  This provides a strong clue to the form of the additional term;
it must be proportional to powers of $\vec{\mud}_i$, and it must vanish
when all the vectors $\vec{\mud}_i$ are parallel.  The missing term,
determined numerically, is the double box planar integral with numerator
$16 (D_s - 2) s_{12} \Bigl( (\mud_p \cdot \mud_q)^2 - \mud_p^2 \, \mud_q^2
\Bigr)$.  We then find for the pure gluon leading-color all-plus helicity
amplitude in \eqn{TwoLoopColor},
$$
A_{4;1;1}^{\LC}(1^+,2^+,3^+,4^+) = A^\P_{1234} + A^\P_{2341} \,,
\equn\label{LeadingGlueResult}
$$
where the `primitive' planar amplitude $A^\P_{1234}$ is analogous
to the one appearing in the $N=4$ super-Yang-Mills case~\cite{BRY},
$$
\eqalign{
A^\P_{1234}
& = i \, { \spb1.2\spb3.4 \over \spa1.2\spa3.4 }
 \biggl\{
 s_{12} \, \I_4^\P \Bigl[ (D_s-2) ( \mud_p^2 \, \mud_q^2 
         + \mud_p^2 \, \mud_{p+q}^2  + \mud_q^2 \, \mud_{p+q}^2 ) 
       + 16 \Bigl( (\mud_p \cdot \mud_q)^2 - \mud_p^2 \, \mud_q^2 \Bigr) 
                    \Bigr](s_{12},s_{23}) \cr
& \hskip 3 cm 
+ 4 (D_s-2) \, \I_4^\bowtie[(\mud_p^2 + \mud_q^2) 
                \, (\mud_p \cdot \mud_q) ] (s_{12}) \cr
& \hskip 3 cm
+ {(D_s-2)^2 \over s_{12}} \, \I_4^\bowtie\Bigl[ 
      \mud_p^2 \, \mud_q^2 \, ( (p+q)^2 + s_{12} ) \Bigr] (s_{12},s_{23})
              \biggr\} \,. \cr}
\equn\label{ppppPlanar}
$$

\subsection{Subleading color}
\label{SubeadingGluonSection}

We may observe that all terms in $A^\LC$ have two-particle cuts
in exactly one channel. If we assume that this property also 
holds for the subleading-color contribution, then performing the sewings
implied by \eqn{TwoCut}, we find that full two-loop amplitude can be 
written in the following form 
$$
{\cal A}_4^\twoloop(1^+,2^+,3^+,4^+) = g^6  
\Bigl[ C^\P_{1234} \, A^\P_{1234} + C^\P_{3421} \, A^\P_{3421}
   + C^\NP_{12;34} \, A^\NP_{12;34} + C^\NP_{34;21} \, A^\NP_{34;21}
    +\ {\cal C}(234) \Bigr] \,,
\equn\label{FColor}
$$
where `$+\ {\cal C}(234)$' instructs one to add the two non-trivial cyclic 
permutations of (2,3,4).  The values of the color coefficients may be
read off directly from \fig{ParentsFigure}.  For example, $C^\P_{1234}$
is the color factor obtained from diagram (a) by dressing each vertex 
with an $\tilde{f}^{abc}$, where 
$$
\tilde{f}^{abc} \equiv i\sqrt{2} f^{abc} = \Tr\bigl( [T^a,T^b] T^c \bigr),
\equn\label{ftildedef}
$$
and dressing each internal line with a $\delta^{ab}$.  Similarly, 
$C^\NP_{12;34}$ is obtained by dressing diagram (b).  The symmetries of 
the color factors can also be read off the diagrams:  
$$
\eqalign{
C^\P_{1234} &= C^\P_{3412} = C^\P_{2143} = C^\P_{4321} \,, \cr
C^\NP_{12;34} &= C^\NP_{21;34} = C^\NP_{12;43} = C^\NP_{21;43} \,. \cr}
\equn\label{Csymmetries}
$$
The corresponding planar and non-planar primitive amplitudes share the
same symmetries with their associated color factors.  Due to these
symmetries, ${\cal A}_4^\twoloop(1^+,2^+,3^+,4^+)$ has the required
total ($S_4$) permutation symmetry, even though only six permutations 
appear explicitly in \eqn{FColor}.

In the amplitude~(\ref{FColor}) no ultraviolet subtraction has been 
performed.  The subtraction in the \MSbar\ scheme is given below in 
\eqn{UVSubtraction}.

The one-channel assumption about the two-particle cuts is non-trivial and
is not expected to hold for general helicity amplitudes.  This property
and the decomposition of \eqn{FColor} are also satisfied by the $N=4$
supersymmetric amplitude~\cite{BRY,TwoLoopGrav}. One may verify the
validity of the assumption {\it a posteriori} by checking the two- and
three-particle cuts, \eqns{TwoCut}{ThreeCut}. Although it might appear
that the expression~(\ref{FColor}) does not assign the proper color
factors to bow-tie integral contributions, the unwanted terms 
(the subleading-color parts of $C^\P$) cancel in the permutation sum.  
The partial amplitudes appearing in the standard color 
decomposition (\ref{TwoLoopColor}) can be expressed in terms of $A^\P$ 
and $A^\NP$.  The equations are identical to those for the $N=4$
supersymmetric case discussed in ref.~\cite{BRY}.

From the point of view of the two-particle cuts, the planar amplitude 
arises from sewing a four-point color-ordered tree to a color-ordered 
one-loop amplitude, where the sewn legs are adjacent in both amplitudes.
In other terms in \eqn{TwoCut}, the permutation of legs on 
the one-loop amplitude ${\cal A}_4^\oneloop$ results in sewing
non-adjacent (diagonally opposite) legs to the tree.  This gives rise
to a primitive amplitude built out of non-planar two-loop integrals,
$$
A^\NP_{12;34}
= i \, { \spb1.2\spb3.4 \over \spa1.2\spa3.4 } \, s_{12} 
\I_4^\NP \Bigl[ (D_s-2) ( \mud_p^2 \, \mud_q^2 
                   + \mud_p^2 \, \mud_{p+q}^2 
                   + \mud_q^2 \, \mud_{p+q}^2 )
+ 16 \Bigl( (\mud_p \cdot \mud_q)^2
           - \mud_p^2 \, \mud_q^2 \Bigr) \Bigr](s_{12},s_{23})\,,
\equn \label{ppppNonPlanar}
$$
where the non-planar double box integral, depicted in
\fig{ParentsFigure}(b), is given by
$$
\eqalign{
\I_4^\NP [{\cal P} (&\mud_i, p,q,k_i)] (s_{12},s_{23}) \cr
 & \equiv \int \! {d^{D} p \over (2\pi)^{D}} \,
        {d^{D} q \over (2\pi)^{D}}\,
{{\cal P} (\mud_i, p, q, k_i) \over p^2\, q^2\, (p+q)^2 \,
         (p-k_1)^2 \,(q-k_2)^2\,
   (p+q+k_3)^2 \, (p+q+k_3+k_4)^2}\,. \cr}
\equn\label{NonPlanarInt}
$$
Note that $A^\NP_{12;34}$ is symmetric under $k_1 \lr k_2$, and 
independently under $k_3 \lr k_4$.  The non-planar amplitude has no
bow-tie contributions; however, the term proportional to 
$\Bigl( (\mud_p \cdot \mud_q)^2 - \mud_p^2 \, \mud_q^2 \Bigr)$
persists.

To verify that the expressions in eqs.~(\ref{FColor}), (\ref{ppppPlanar})
and (\ref{ppppNonPlanar}) give the correct full gluon amplitude, we have
evaluated all the independent cuts.  Consider, for example, the
leading-color planar contribution proportional to the color factor $N_c^2
\Tr[T^{a_1} T^{a_2} T^{a_3} T^{a_4}]$, which is given by
\eqn{LeadingGlueResult}.  For this contribution, the $s_{12}$-channel
three-particle cut integrand should be equal to the product of
color-ordered five-gluon tree-level partial amplitudes,
$$
A_5^{\tree}(1^+, 2^+, -\ell_3^\rho, -\ell_2^\nu, -\ell_1^\mu) \times
P_{\mu\alpha}(\ell_1,r) P_{\nu\beta}(\ell_2,r) 
P_{\rho\gamma}(\ell_3,r)  \times
A_5^{\tree}(\ell_1^\alpha, \ell_2^\beta, \ell_3^\gamma, 3^+, 4^+) \,.
\equn \label{TreeProduct}
$$
We have verified numerically that the product of tree
amplitudes~(\ref{TreeProduct}) is equal to the three-particle cut of
the leading-color expression (\ref{LeadingGlueResult}), given by
$$
\eqalign{
& i {\spb1.2\spb3.4 \over \spa1.2 \spa3.4}  \Bigl[
    (D_s-2) (\mud_1^2\mud_2^2 + \mud_1^2\mud_3^2 + \mud_2^2\mud_3^2)
  + 16 \Bigl((\mud_1\cdot\mud_3)^2 - \mud_1^2 \mud_3^2 \Bigr) \Bigr]\cr
& \hskip .3 cm  \times
 \biggl[  {s_{23} \over (\ell_1-k_1)^2 (\ell_3-k_2)^2 (\ell_3+k_3)^2 
        (\ell_1+k_4)^2}
        +{s_{12} \over (\ell_1+\ell_2)^2 (\ell_3-k_2)^2 (\ell_2+\ell_3)^2 
       (\ell_1+k_4)^2} \cr 
& \hskip 3.5 cm 
        +{s_{12} \over (\ell_1+\ell_2)^2 (\ell_3+k_3)^2 (\ell_2+\ell_3)^2 
       (\ell_1-k_1)^2} \biggr] \,. \cr}
\equn\label{PureGluonThreeCut}
$$
The three different terms in the last factor in \eqn{PureGluonThreeCut}
originate from the two cut planar double box integrals in the same way as
in the scalar example~(\ref{AThreeCut}), as illustrated in
\fig{TripleCutFigure}.  

As a second example of a $s_{12}$-channel three-particle cut, but one that
is sensitive to non-planar (subleading-color) contributions, we evaluate
the product
$$
A_5^{\tree}(1^+, -\ell_3^\rho, -\ell_2^\nu, 2^+, -\ell_1^\mu) \times
P_{\mu\alpha}(\ell_1,r) P_{\nu\beta}(\ell_2,r) 
P_{\rho\gamma}(\ell_3,r)  \times
A_5^{\tree}(\ell_1^\alpha, \ell_2^\beta, \ell_3^\gamma, 3^+, 4^+) 
\equn \label{SubleadingTreeProduct}
$$
numerically and find that it equals
$$
\eqalign{
& i {\spb1.2\spb3.4 \over \spa1.2 \spa3.4}  \Bigl[
    (D_s-2) (\mud_1^2\mud_2^2 + \mud_1^2\mud_3^2 + \mud_2^2\mud_3^2)
  + 16 \Bigl((\mud_1\cdot\mud_3)^2 - \mud_1^2 \mud_3^2 \Bigr) \Bigr]\cr
& \hskip .3 cm  \times
 \biggl[ - { s_{13} \over 
       (\ell_3-k_1)^2 (\ell_1-k_2)^2 (\ell_3+k_3)^2 (\ell_1+k_4)^2 }
         - { s_{13} \over 
       (\ell_3-k_1)^2 (\ell_2-k_2)^2 (\ell_3+k_3)^2 (\ell_1+k_4)^2 } \cr
& \hskip 1.5 cm 
         + { s_{23} \over 
       (\ell_1-k_1)^2 (\ell_2-k_2)^2 (\ell_3+k_3)^2 (\ell_1+k_4)^2 }
         + { s_{12} \over 
       (\ell_3-k_1)^2 (\ell_2-k_2)^2 (\ell_2+\ell_3)^2 (\ell_1+k_4)^2 } \cr
& \hskip 1.5 cm 
         - { s_{12} \over 
       (\ell_1-k_1)^2 (\ell_2-k_2)^2 (\ell_3+k_3)^2 (\ell_1+\ell_2)^2 }
         - { s_{12} \over 
       (\ell_1-k_1)^2 (\ell_2+\ell_3)^2 (\ell_3+k_3)^2 (\ell_1+\ell_2)^2 }\cr
& \hskip 1.5 cm 
         - { s_{12} \over 
       (\ell_1-k_2)^2 (\ell_2+\ell_3)^2 (\ell_3+k_3)^2 (\ell_1+\ell_2)^2 }
\biggr] \,. \cr}
\equn\label{PureGluonSubleadingThreeCut}
$$
The latter candidate expression is found by extracting from \eqn{FColor}
the seven terms with the correct color ordering to contribute to the
cut~(\ref{SubleadingTreeProduct}).  (Three different planar double box
integrals and three different non-planar ones appear; one of the
non-planar integrals can be cut two separate ways.  Minus signs are due to
the antisymmetry of the structure constants in the color factors.)  More
generally, we have verified that the amplitude~(\ref{FColor}) has the
correct $D$-dimensional two- and three-particle cuts in all channels,
proving it to be correct.

We have also shown that the double two-particle and three-particle cuts of
the Feynman gauge Feynman diagrams, including the ghosts, match the cuts
of our expressions.  In performing this consistency check there is no need
to include physical state projectors on the intermediate gluon lines; the
ghosts automatically cancel the unwanted longitudinal modes.  We carried
out the comparison numerically for specific values of the (cut) loop
momenta.  This provides a non-trivial check that our expressions are in
one-to-one correspondence with the results that one would obtain via a
direct evaluation of the Feynman diagrams.

In the appendix we perform the final step in a closed-form evaluation
of the two-loop amplitude:  We present the values of the integrals
appearing in the amplitudes, expanded in $\e$ through $\Ord(\e^0)$ and
expressed in terms of polylogarithms.  The bow-tie
integrals are rather simple to evaluate since they are just products
of one-loop integrals. The planar and non-planar double box integrals
are much more difficult to obtain, because they have poles up to order
$1/\e^2$ and an intricate analytic structure (especially the non-planar 
cases).


\section{Divergences}
\label{DivergenceSection}

In this section we compare the divergence structure of the all-plus
helicity amplitude (\ref{FColor}) against the one expected from general
principles.  Catani has previously presented a universal factorization
formula for the infrared divergent parts of dimensionally regulated,
renormalized two-loop amplitudes~\cite{CataniDiv}.  In the all-plus case,
the divergence structure is relatively simple, due to the vanishing of the
corresponding tree-level helicity amplitude; both the ultraviolet and
infrared divergences are essentially the same ones encountered at one
loop~\cite{KunsztSingular}.  Nevertheless, we find it convenient to use
Catani's formalism, because of its more general applicability.

In the color space operator language used by Catani, the infrared
divergences of a renormalized two-loop amplitude are,
$$
\eqalign{
| \cm_n^{(2)}(\mu_R^2; \{p\}) \ra_{\RS} &= 
{\bom I}^{(1)}(\eps, \mu_R^2; \{p\}) 
\; | \cm_n^{(1)}(\mu_R^2; \{p\}) \ra_{\RS} \cr
& \hskip 1 cm 
+ {\bom I}^{(2)}_{\RS}(\eps, \mu_R^2; \{p\}) \; 
  | \cm_n^{(0)}(\mu_R^2; \{p\}) \ra_{\RS}
+ \hbox{finite} \,, \cr}
\equn\label{TwoLoopCatani}
$$
where $|\cm_n^{(L)}(\mu_R^2; \{p\}) \ra_{\RS}$ is a color space
vector representing the renormalized $L$ loop amplitude.  The
subscript $\RS$ stands for the choice of renormalization scheme,
and $\mu_R$ is the renormalization scale (which we have set to
unity elsewhere in the paper).  These color space vectors give the
amplitudes via,
$$
{\cal A}_n(1^{a_1},\dots,n^{a_n}) \equiv
\la a_1,\dots,a_n \,
| \, \cm_n(p_1,\ldots,p_n)\ra \,,
\equn
$$
where the $a_i$ are color indices.  The divergences of ${\cal A}_n$
are encoded in the color operators 
${\bom I}^{(1)}(\eps,\mu_R^2;\{p\})$ and 
${\bom I}^{(2)}(\eps,\mu_R^2;\{p\})$.  For the all-plus helicity case, 
the tree amplitude vanishes so ${\bom I}^{(2)}(\eps,\mu_R^2;\{p\})$ does
not enter.  The operator ${\bom I}^{(1)}(\eps,\mu_R^2;\{p\})$ is given 
by
$$
{\bom I}^{(1)}(\eps,\mu_R^2;\{p\}) =  \frac{1}{2}
{2\over (4\pi)^{2-\eps}} {1\over\Gamma(1-\eps)} \sum_{i=1}^n
\, \sum_{j \neq i}^n \, 
{\bom T}_i \cdot {\bom T}_j \biggl[ {1\over \eps^2}
\biggl( \frac{\mu_R^2 e^{-i\lambda_{ij} \pi}}{2 p_i\cdot p_j} \biggr)^{\eps}
 + {\gamma_i \over {\bom T}_i^2 } \, {1\over \eps} \biggr]
 \,,
\equn\label{CataniGeneral}
$$
where $\lambda_{ij}=+1$ if $i$ and $j$ are both incoming or
outgoing partons and $\lambda_{ij}=0$ otherwise. The color charge
${\bom T}_i = \{T^a_i\}$ is a vector with respect to the generator label
$a$, and an $SU(N_c)$ matrix with respect to the color indices of
parton $i$.  For the adjoint representation $T^a_{cb} = i f^{cab}$,
and so ${\bom T}_i^2 = C_A = N_c$.  Also, for external gluons in a pure
glue theory, $\gamma_g = \textstyle{11\over6} C_A$.
(We have not included an overall factor of $e^{-\eps\psi(1)}$ appearing 
in Catani's expression, but have included a factor of $2/(4\pi)^{2-\eps}$ 
because of differing overall normalization conventions.) 

Specializing ${\bom I}^{(1)}(\eps,\mu_R^2;\{p\})$ to the pure gluon case,
and then rewriting the divergent terms in \eqn{TwoLoopCatani} in our 
notation gives,
$$
\eqalign{
 {\cal A}_n^{\twoloop,\, \rm ren.}(1^{a_1}, 2^{a_2}, \ldots, n^{a_n})
  \Bigr|_{\rm div.} & = 
  \sum_{i<j}^n {\cal A}_n^{(i,j),\twoloop,\, \rm ren.}(1,2,3,\ldots,n)  
                                      \Bigr|_{\rm div.} \,, \cr} 
\equn\label{CataniPureGlueA}
$$
where 
$$
\eqalign{
{\cal A}_n^{(i,j),\twoloop,\, \rm ren.}(1,2,3,\ldots,n) \Bigr|_{\rm div.}
&\equiv 
- g^2 \cg \, \tilde{f}^{ b_i a_i c} \tilde{f}^{c a_j b_j} 
 \biggl[ {1\over \eps^2} (- s_{ij})^{-\eps} 
     + \, {11\over 6}\, {1\over \eps} \biggr]\cr
& \hskip 1. cm 
\times {\cal A}_n^\oneloop(1^{a_1}, 2^{a_2}, \ldots, i^{b_i}, \ldots, 
        j^{b_j}, \ldots, n^{a_n}) \,, \cr}
\equn\label{CataniPureGlueB}
$$
and $\tilde{f}^{abc}$ is defined in \eqn{ftildedef}.  In
converting the overall normalization we used
$$
\eqalign{
\cg \equiv {1\over (4\pi)^{2-\eps}} 
{\Gamma(1+\eps) \Gamma^2(1-\eps) \over \Gamma(1-2\eps)}  \cr
= {1\over (4\pi)^{2-\eps}} {1 \over \Gamma(1-\eps)}   + \Ord(\eps^3)\,.\cr}
\equn\label{cgdef}
$$

Equation~(\ref{CataniPureGlueA}) holds for every pure gluon helicity 
amplitude of the form ${\cal A}_n(\pm,+,+,\ldots,+)$;
such amplitudes vanish at tree-level due to supersymmetry Ward 
identities~\cite{SWI}.  (Although the pure gluon theory is not
supersymmetric, its tree amplitudes obey supersymmetry identities because
the gluons' fermionic superpartners cannot contribute at tree level.)
Specializing now to the four-point case, we have
$$
{\cal A}_4^{\twoloop,\,\rm ren.}(1,2,3,4)\Bigr|_{\rm div.}
\equiv \sum_{i<j}^4 {\cal A}_4^{(i,j),\twoloop,\, \rm ren.}(1,2,3,4) 
  \Bigr|_{\rm div.}\,.
\equn\label{DivergenceSum}
$$
We shall evaluate the term in \eqn{DivergenceSum} with $(i,j) = (1,2)$,
$$
\eqalign{
{\cal A}_4^{(1,2),\,\twoloop, \,\rm ren.}(1^{a_1}, 2^{a_2}, 3^{a_3}, 4^{a_3}) 
\Bigr|_{\rm div.} & = 
- g^2 \cg \biggl[ {1\over \eps^2}( -s_{12})^{-\eps} 
      + {11\over 6} {1\over \eps} \biggr]
\,  \tilde{f}^{b_1 a_1 c} \tilde{f}^{c a_2 b_2} \cr
& \hskip 3 cm
\times {\cal A}^\oneloop_4(1^{b_1}, 2^{b_2}, 3^{a_3}, 4^{a_4}) \,, \cr}
\equn\label{FourPointDiv}
$$
and then obtain the other five cases by relabeling $i$ and $j$.   
Note that we count $(i,j)=(3,4)$ as being distinct from $(i,j)=(1,2)$ 
in the sum~(\ref{DivergenceSum}).

The one-loop amplitudes may be decomposed in terms of $SU(N_c)$ 
structure constants~\cite{LanceColor} (see also eq. (2.30) of 
ref.~\cite{TwoLoopGrav}),
$$
\eqalign{
{\cal A}_4^{{\oneloop}} (1,2,3,4) & =
   g^4 \Bigl[ C_{1234}^\oneloop\, A_4^{\oneloop}(1,2,3,4)
         + C_{1243}^\oneloop\, A_4^{\oneloop}(1,2,4,3) \cr
& \hskip 2 cm 
         + C_{1423}^\oneloop\, A_4^{\oneloop}(1,4,2,3) \Bigr] \,, \cr}
\equn\label{OneLoopYMResult}
$$
where $C_{1234}^\oneloop$ is the color factor obtained by dressing
each vertex in the one-loop box diagram in \fig{OneLoopColorFigure}
with a structure constant $\tilde{f}^{abc}$, and dressing each bond
between vertices with a $\delta^{ab}$.  In this decomposition the
$A_4^{\oneloop}$ are free of all group theory factors, and are invariant
under cyclic permutations of their four arguments.

For the all-plus helicity case, the primitive amplitudes $A_4^{\oneloop}$
are proportional to a one-loop box integral~\cite{DimShift}, 
$$
 A_4^\oneloop(1^+,2^+,3^+,4^+) =  (D_s -2)\,
 {\spb1.2 \spb3.4\over \spa1.2 \spa3.4} \ 
\I_4^\oneloop[\mud_p^4](s_{12},s_{23})\,,
\equn
$$
with
$$
\I_4^\oneloop[\mud_p^4](s_{12},s_{23}) 
\equiv \int {d^D p \over (2\pi)^{D}} \, 
{\mud_p^4 \over p^2 (p-k_1)^2 (p-k_1-k_2)^2 (p+k_4)^2} \,.
\equn\label{OneLoopIntDef}
$$

%
\begin{figure}[ht]
\begin{center}
\begin{picture}(70,70)(0,0)
\Text(8,10)[r]{1} \Text(8,60)[r]{2}
\Line(10,60)(20,50)  \Line(50,20)(60,10) 
\Line(10,10)(20,20) \Line(50,50)(60,60)
\Line(20,20)(50,20) \Line(50,20)(50,50) 
\Line(50,50)(20,50) \Line(20,50)(20,20)
\Text(62,60)[l]{3} \Text(62,10)[l]{4}
\end{picture}
\caption[]{
\label{OneLoopColorFigure}
\small $C_{1234}^\oneloop$ is given by dressing each vertex in the 
figure with $\tilde{f}^{abc}$, and dressing each bond between
vertices with a $\delta^{ab}$.}
\end{center}
\end{figure}
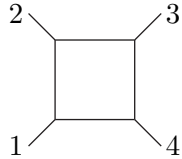

%
\begin{figure}[ht]
\begin{center}
\begin{picture}(380,70)(0,0)
\Text(38,15)[r]{1} \Text(38,55)[r]{2}
\Line(60,15)(50,25) \Line(40,15)(50,25)
\Line(60,55)(50,45) \Line(40,55)(50,45)    \Line(50,25)(50,45)
\Text(63,55)[l]{$b_2$} \Text(63,15)[l]{$b_1$}
\Text(90,35)[c]{\Large $\times$}
\Line(110,10)(115,10) \Line(110,10)(110,60) \Line(110,60)(115,60) 
\Text(133,15)[r]{$b_1$} \Text(133,55)[r]{$b_2$}
\Line(135,55)(145,45)  \Line(165,25)(175,15) 
\Line(135,15)(145,25) \Line(165,45)(175,55)
\Line(145,25)(165,25) \Line(165,25)(165,45) 
\Line(165,45)(145,45) \Line(145,45)(145,25)
\Text(178,55)[l]{3} \Text(178,15)[l]{4}
\Text(200,35)[c]{\Large +}
\Text(223,15)[r]{$b_1$} \Text(223,55)[r]{$b_2$}
\Line(225,55)(235,45)  \Line(255,25)(265,15) 
\Line(225,15)(235,25) \Line(255,45)(265,55)
\Line(235,25)(255,25) \Line(255,25)(255,45) 
\Line(255,45)(235,45) \Line(235,45)(235,25)
\Text(268,55)[l]{4} \Text(268,15)[l]{3}
\Text(290,35)[c]{\Large +}
\Text(313,15)[r]{$b_1$} \Text(313,55)[r]{4}
\Line(315,55)(325,45)  \Line(345,25)(355,15) 
\Line(315,15)(325,25) \Line(345,45)(355,55)
\Line(325,25)(345,25) \Line(345,25)(345,45) 
\Line(345,45)(325,45) \Line(325,45)(325,25)
\Text(358,55)[l]{$b_2$} \Text(358,15)[l]{3}
\Line(370,10)(375,10) \Line(375,10)(375,60) \Line(370,60)(375,60) 
\end{picture}
\caption[]{
\label{DivergencesFigure}
\small A schematic version of the color factors in
\eqn{FinalDivergence}. The color factors are given by dressing each
vertex with $\tilde{f}^{abc}$, and dressing each bond between vertices
with a $\delta^{ab}$.}
\end{center}
\end{figure}
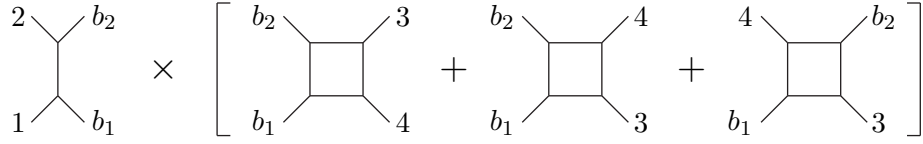

Using \eqn{OneLoopYMResult} the divergence (\ref{FourPointDiv}) is
$$
\eqalign{
{\cal A}^{(1,2),\,\twoloop,\, \rm ren.}_4(1,2,3,4) \Bigr|_{\rm div.} & = 
- g^6 \cg
 \biggl[ {1\over \eps^2} (-s_{12})^{-\eps} 
 + {11\over 6} {1\over \eps} \biggr]
\, \tilde{f}^{b_1 a_1 c} \tilde{f}^{c a_2 b_2} 
 \Bigl( C_{b_1b_234}^\oneloop\, A_4^{\oneloop}(1,2,3,4) \cr
& \hskip 2.5 cm
            + C_{b_1b_243}^\oneloop\, A_4^{\oneloop}(1,2,4,3) 
            + C_{b_14b_23}^\oneloop\, A_4^{\oneloop}(1,4,2,3)\Bigr)\,.\cr}
\equn\label{DivergenceAlmostFinal}
$$
We display a schematic version of the above color factors
in \fig{DivergencesFigure}.  Anticipating our comparison to the
divergences appearing in the two-loop amplitude (\ref{FColor}), we
observe that the one-loop color coefficients can be related to the
two-loop ones via,
$$
\eqalign{
& \tilde{f}^{b_1 a_1 c} \tilde{f}^{c a_2 b_2} C_{b_1b_234}^\oneloop 
                 = C^\P_{1234}\,, \cr
& \tilde{f}^{b_1 a_1 c} \tilde{f}^{c a_2 b_2} C_{b_1b_243}^\oneloop 
                 = C^\P_{1243}\,, \cr
& \tilde{f}^{b_1 a_1 c} \tilde{f}^{c a_2 b_2} C_{b_14b_23}^\oneloop 
                 = C^\NP_{34;12}\,. \cr}
\equn
$$
Using these relations in \eqn{FourPointDiv} we have 
$$
\eqalign{
{\cal A}^{(1,2),\,\twoloop,\,\rm ren.}_4(1,2,3,4) \Bigr|_{\rm div.} & = 
- g^6 \cg \biggl[ {1\over \eps^2} (- s_{12})^{-\eps} 
                + {11\over 6} {1\over \eps} \biggr] \, 
 \Bigl( C^\P_{1234} \, A_4^{\oneloop}(1,2,3,4) \cr
& \hskip 2 cm
            + C^\P_{1243} \, A_4^{\oneloop}(1,2,4,3) 
            + C^\NP_{34;12}\, A_4^{\oneloop}(1,4,2,3) 
               \Bigr) \,, \cr}
\equn\label{DivColor}
$$
so the color factors are the same ones appearing in the two-loop
amplitude.  Inserting \eqn{DivColor} into \eqn{DivergenceSum},
summing over $i$ and $j$, and making use of \eqn{Csymmetries}, 
we obtain the predicted divergence,
$$
\eqalign{
&{\cal A}_4^{\twoloop,\, \rm ren.}(1,2,3,4)\Bigr|_{\rm div.} \cr
& \hskip 1 cm
= -g^6 \cg \biggl[ {1 \over \eps^2} 
 (-s_{12})^{-\eps} + {11\over 6} {1 \over \eps} \biggr]
      \Bigl( 2 C^\P_{1234} A_{4}^\oneloop(1,2,3,4) 
           + 2 C^\P_{3421} A_{4}^\oneloop(3,4,2,1) \cr
& \hskip 5.7 cm 
           + C^\NP_{12;34} \, A_{4}^\oneloop(1,3,2,4) 
           + C^\NP_{34;21} \, A_{4}^\oneloop(3,2,4,1) \Bigr)
           +  {\cal C}(234)\,. \cr}
\equn\label{FinalDivergence}
$$

Now we compare \eqn{FinalDivergence} to the divergences found in the 
two-loop all-plus helicity amplitude (\ref{FColor}), using results
for the two-loop integrals given in the appendix.  All of the integrals 
are finite except for $\I_4^\P[\mud_{p+q}^2 \mud_q^2]$,
$\I_4^\P[\mud_{p+q}^2 \mud_p^2]$ and $\I_4^\NP[\mud_p^2 \mud_q^2]$.
For \eqn{FinalDivergence} to be recovered, it is crucial that 
the divergent parts of these integrals are proportional to the same 
one-loop box integral~(\ref{OneLoopIntDef}) encountered in 
$A_4^\oneloop$:
$$
\eqalign{
  \I_4^\P[\mud_{p+q}^2 \mud_q^2](s_{12},s_{23}) \Bigr|_{\rm div.}
= \I_4^\P[\mud_{p+q}^2 \mud_p^2](s_{12},s_{23}) \Bigr|_{\rm div.}
&= - i \, \cg \, {1 \over \e^2} \, (-s_{12})^{-1-\e}
     \, \I_4^\oneloop[\mud_p^4](s_{12},s_{23}) \,, \cr 
  \I_4^\NP[\mud_p^2 \mud_q^2](s_{12},s_{23}) \Bigr|_{\rm div.}
&= - i \, \cg \, {1 \over \e^2} \, (-s_{12})^{-1-\e}
     \, \I_4^\oneloop[\mud_p^4](s_{13},s_{23}) \,, \cr }
\equn\label{FactorizeSingIntegrals}
$$
as can be seen from \eqns{PlanarMinusSing}{NonPlanarMinusSing}.

This factorization of the singular behavior of the integrals can be
understood heuristically.  The $1/\e^2$ poles come from loop momenta that
are simultaneously soft and collinear with two adjacent external legs, so
that three consecutive propagators can go on shell.  Because the external
momenta are strictly four-dimensional, the singular region of loop momenta
is also four-dimensional.  However, a factor of $\mud_p^2$ in the
numerator suppresses the region where $p$ becomes four-dimensional.  Thus
in $\I_4^\P[\mud_{p+q}^2 \mud_q^2]$, for example, the region with
$\mud_p^2 \approx 0$ and $p^2 \approx 0$ dominates the integral, while $q$
and $p+q$ remain off shell.  The space-time picture of the dominant
integration region corresponds to shrinking the rightmost box of the
planar double box in \fig{ParentsFigure}(a) to a small size.  The
kinematics of that inner box integral is hardly affected by the soft
momentum $p$, so it is essentially the same as the external kinematics,
and it can be factored out of the outer triangle integral that remains.
The triangle integral gives the factor of 
$-i \, \cg \, (-s_{12})^{-1-\e} / \e^2$ `dressing' the box integral.  

Apart from those in \eqn{FactorizeSingIntegrals}, all the other integrals
appearing in the two-loop all-plus amplitude are finite in the limit 
$\e\to0$.  For these integrals, whenever there is a loop momentum from
the set $\{ p, \, q, \, p+q \}$ whose four-dimensional limit is not 
suppressed by a numerator factor of $\mud_i^2$, then that loop momentum 
has at most two propagators containing it, not three.  This property 
prevents an infrared divergence from occurring.

Using eqs.~(\ref{FactorizeSingIntegrals}), (\ref{ppppPlanar}) and 
(\ref{ppppNonPlanar}), and the total symmetry of the prefactor 
$\spb1.2\spb3.4/(\spa1.2\spa3.4)$, the divergences of the
bare (unrenormalized) primitive amplitudes are
$$
\eqalign{
A^\P_{1234}\Bigr|_{\rm div.} & = - 2 \, \cg \, 
    {1 \over \eps^2} (-s_{12})^{-\eps}
         A_4^\oneloop(1^+,2^+,3^+,4^+) \,, \cr
A^\NP_{12;34}\Bigr|_{\rm div.} & = - \cg \, 
     {1 \over \eps^2} (-s_{12})^{-\eps}
         A_4^\oneloop(1^+,3^+,2^+,4^+) \,. \cr}
\equn\label{FactorizeSingAmplitudes}
$$
Inserting these expressions into the bare full amplitude (\ref{FColor}) 
then yields
$$
\eqalign{
{\cal A}_4^\twoloop(1,2, 3,4)\Bigr|_{\rm div.} & =  - g^6 \cg
\biggl[ {2 \over \eps^2} (-s_{12})^{-\eps}
            C^\P_{1234} A_{4}^\oneloop(1,2,3,4) +
 {2 \over \eps^2} (-s_{12})^{-\eps}
            C^\P_{3421} A_{4}^\oneloop(3,4,2,1) \cr
& \hskip 1 cm 
+ {1 \over \eps^2} (-s_{12})^{-\eps} 
            C^\NP_{12;34} \, A_{4}^\oneloop(1,3,2,4) 
+ {1 \over \eps^2} (-s_{12})^{-\eps} 
            C^\NP_{34;21} \, A_{4}^\oneloop(3,2,4,1) \biggr] \cr
& \hskip 3 cm 
           +  {\cal C}(234) \,. \cr}
\equn\label{FullDiverge}
$$

The expression~(\ref{FullDiverge}) already agrees with 
\eqn{FinalDivergence} at the level of the $(-s_{ij})^{-\e}/\e^2$ poles.  
To check the agreement at the $1/\e$ level, we must first renormalize our
two-loop amplitude in the \MSbar\ scheme.  The \MSbar\ counterterm to 
be subtracted from ${\cal A}_4^\twoloop(1,2,3,4)$ is 
$$
\eqalign{
 \hbox{C.T.} &= 4 \, g^2\, \cg {11 N_c \over 6} {1\over\eps} 
    {\cal A}_4^\oneloop(1^+,2^+,3^+,4^+)\,, \cr
& = 4 N_c \, g^6\, \cg {11\over6} {1\over\e} 
 \Bigl[ C_{1234}^\oneloop\, A_4^{\oneloop}(1,2,3,4)
      + C_{3124}^\oneloop\, A_4^{\oneloop}(3,1,2,4)
      + C_{2314}^\oneloop\, A_4^{\oneloop}(2,3,1,4)  \Bigr] . \cr }
\equn\label{CT1}
$$
The relative simplicity of the ultraviolet subtraction term is due to 
the vanishing of the corresponding tree-level helicity amplitude;
the counterterm has the same structure encountered in other helicity
configurations at one loop, up to the overall factor and the
replacement of a tree amplitude with the one-loop amplitude.  
(At two loops, an additional $1/\eps^2$ subtraction is required for
helicity configurations with non-vanishing tree amplitudes.
Additional subtraction terms are necessary for amplitudes including 
fermions or scalars as well.)

Finally, we use the color conservation identity 
$\sum_i {\bom T}_i = 0$~\cite{Catani}, which implies that
$$
n \, N_c\  | \cm_n \ra
= \sum_{i=1}^n {\bom T}_i^2\  | \cm_n \ra
= - \sum_{j \not = i}^n {\bom T}_i \cdot  {\bom T}_j\  | \cm_n \ra \,,
\equn\label{ColorIdentity}
$$
when all external particles are in the adjoint representation.
Replacing the $4N_c$ in \eqn{CT1} with the help of \eqn{ColorIdentity},
the color structure of the \MSbar\ counterterm (and its dependence
on $A_4^\oneloop$) can be put in exactly the same form as Catani's 
expression for the divergent terms, eqs.~(\ref{TwoLoopCatani}), 
(\ref{CataniGeneral}) and (\ref{CataniPureGlueA}).
We can then process it into the form~(\ref{FinalDivergence}) in just the 
same way, obtaining
$$
\eqalign{
 \hbox{C.T.} & = g^6\, \cg \biggl[ 
  {11\over3}\, {1\over\eps} C^\P_{1234} A_{4}^\oneloop(1,2,3,4)
+ {11\over3}\, {1\over\eps} C^\P_{3421} A_{4}^\oneloop(3,4,2,1) \cr
& \hskip 1 cm 
+ {11\over6}\,  {1 \over \eps} 
            C^\NP_{12;34} \, A_4^\oneloop(1,3,2,4)
+ {11\over6}\,  {1 \over \eps} 
            C^\NP_{34;21} \, A_4^\oneloop(3,2,4,1) \biggr] 
           +  {\cal C}(234) \,. \cr}
\equn\label{UVSubtraction}
$$
After including the ultraviolet subtraction~(\ref{UVSubtraction}), the 
infrared divergences of our renormalized amplitude agree perfectly with 
the expected form~(\ref{FinalDivergence}).


\section{Conclusions}

In this paper we presented the pure glue contribution to the identical
helicity two-loop four-gluon amplitude in QCD.  We calculated the
amplitude using its analytic properties in $D$ dimensions.  More
generally, any amplitude in a massless theory can in principle be obtained
in this way.  Although QCD is not supersymmetric, we also used
supersymmetry as a helpful guide in constructing the pure gluon amplitude
from amplitudes with scalar loops.  Cases with quarks will be discussed
elsewhere.

We expect the type of methods used here to be useful for obtaining 
more general two-loop amplitudes relevant for high energy processes.
Although much more remains to be done, we are hopeful that
calculations of next-to-next-to-leading-order multi-jet cross 
sections will become possible.


\vskip .3 cm 
\noindent
{\bf Acknowledgments}

We thank R. Blankenbecler, H. Haber, K. Melnikov and M. Peskin for useful
discussions regarding two-loop integrals, and A. Sabio Vera for finding
some typographical errors in the original version of this article.


\appendix
\section{Integrals}

In this appendix we evaluate the integrals appearing in the all-plus 
helicity amplitude, in an expansion in $\e$ through $\Ord(\eps^0)$.  

The planar bow-tie integrals are rather simple to evaluate because they
are products of one-loop integrals,
$$
\eqalign{
\I_4^\bowtie[\mud_p^2 \mud_q^2](s) &= - {1\over 4} \, 
   {1\over (4 \pi)^4}\,, \cr
\I_4^\bowtie[\mud_p^2 \mud_q^2 (p+q)^2](s,t) 
&= - {1\over 36} {1\over (4\pi)^4} (t-4s) \,, \cr
\I_4^\bowtie[\mud_p^2 \, (\mud_p \cdot \mud_q)](s) &= 0\,. \cr}
\equn\label{BowTieInts}
$$
Two of the planar double box integrals and one of the non-planar ones
vanish through $\Ord(\eps^0)$, 
$$
\eqalign{
\I_4^\P[(\mud_p \cdot \mud_q)^2](s,t) &= 0 \,, \cr
\I_4^\P[\mud_p^2 \mud_q^2](s,t) &= 0 \,, \cr
\I_4^\NP[ (\mud_p \cdot \mud_q)^2 - \mud_p^2 \mud_q^2 ](s,t) &= 0 \,.  \cr}
$$

The non-vanishing planar double box integral, evaluated in its Euclidean 
region where $s,t<0$ and $u>0$, is,
$$
\eqalign{
& \I_4^\P[\mud_{p+q}^2 \mud_q^2](s,t) 
= \I_4^\P[\mud_{p+q}^2 \mud_p^2](s,t) \cr
& \hskip 0.3 cm
= \ctg \, (-s)^{-1-2\e} \, {1\over6} 
 \biggl\{ -{1\over\e^2} 
   + {1\over\e}
  \biggl[ {1\over2} { \chi (\ln^2\chi + \pi^2) \over (1+\chi)^2 }
        + {\chi\ln\chi\over 1+\chi} - {8\over3} \biggr] \cr
&\hskip 1.3 cm 
   + { \chi \over (1+\chi)^2 } \biggl[ 
         - \Li_3(-\chi) + \zeta(3) + \ln\chi \, \Li_2(-\chi) 
         + \Bigl( {1\over2} \ln(1+\chi) - {2\over3} \ln\chi 
                - {1\over2} \chi + {1\over3} \Bigr) (\ln^2\chi + \pi^2)
         \cr
&\hskip 1.3 cm 
         + \Bigl( {8\over3} (1+\chi) - {\pi^2\over6} \Bigr) \ln\chi \biggr] 
       - {\pi^2\over6} { 2-5\chi \over 1+\chi } - {52 \over 9} \biggr\} \,,
   \cr}
\equn\label{PlanarInts}
$$
where $\chi = t/s$ and
$$
 \ctg \equiv { 1 \over (4\pi)^{4-2\e} } 
    { \Gamma(1+2\e) \Gamma^3(1-\e) \over \Gamma(1-3\e) } \,.
\equn\label{ctgdef}
$$
The polylogarithms are~\cite{Lewin}
$$
\Li_2(x) = -\int_0^x dt \, {\ln(1-t) \over t}\,, \hskip 2 cm
\Li_3(x) = \int_0^x dt \, {\Li_2(t) \over t}\,,
\equn\label{PolylogDefn}
$$
and $\zeta(3) = \Li_3(1) = 1.202\ldots$.
\Eqn{PlanarInts} was originally found by direct integration over 
Feynman parameters, after performing certain subtractions to allow
expansions of integrands in $\e$.  We have also checked that the same 
result can be obtained via the general tensor integral reduction method 
of Smirnov and Veretin~\cite{Smirnov}.

The remaining planar integrals in the full amplitude~(\ref{FColor}) 
may be obtained from \eqn{PlanarInts} by relabeling the legs, which
induces transformations on $\chi$, for example,
$$
\eqalign{
 s \lr t \hskip1cm  &\Leftrightarrow \hskip1cm \chi \lr {1\over\chi}\,, \cr 
 t \lr u \hskip1cm  &\Leftrightarrow \hskip1cm \chi \lr -1-\chi, \cr 
 s \lr u \hskip1cm  &\Leftrightarrow \hskip1cm 
                            \chi \lr {-\chi \over 1+\chi}\,, \cr }
\equn\label{chiPairSwaps}
$$
using $s+t+u=0$.

For two reasons, it is useful to represent the poles in $\e$ of
\eqn{PlanarInts}, and of the non-planar integral $\I_4^\NP[ \mud_p^2
\mud_q^2 ]$, in terms of the one-loop box integral $\I_4^\oneloop[
\mud_p^4 ]$ defined in \eqn{OneLoopIntDef}.  First, the divergence
structure of the two-loop all-plus helicity amplitude becomes more
transparent, as discussed in section~\ref{DivergenceSection}, given
that $\I_4^\oneloop[ \mud_p^4 ]$ appears in the corresponding one-loop
amplitude.  Second, the description of the finite ($\Ord(\e^0)$)
remainder terms is simplified, especially regarding their values in
different kinematic regions.  The heuristic reason why the singular
parts of the two-loop integrals are proportional to $\I_4^\oneloop[
\mud_p^4 ]$ was given in section~\ref{DivergenceSection}.

The one-loop box integral~(\ref{OneLoopIntDef}) can be performed by first
integrating over the $(-2\e)$-dimensional components of $p$, which results
in an integral of the scalar type but in $8-2\e$ dimensions, not $4-2\e$.
A standard one-loop `dimension-shifting' identity allows one to trade this
integral for a quantity ${\cal L}(s,t)$ which is essentially the scalar
integral in $6-2\e$ dimensions~\cite{OurIntegrals}.  This integral has
neither ultraviolet nor infrared divergences, which makes it
straightforward to expand in $\e$.

Concretely, we have
$$
\eqalign{
 \I_4^\oneloop[ \mud_p^4 ](s,t) 
&=  -\e(1-\e) \, (4\pi)^2 \, \I_4^{\oneloop,\, D=8-2\e}[1](s,t) \cr
&= { -\e(1-\e) \over (1-2\e)(3-2\e) } \biggl[
  - {1\over4} { \chi \, {\cal L}(s,t) \over (1+\chi)^2 }
  + {1\over2} { i \, \cg \, (-s)^{-\e} \over \e(1-\e) } 
              { 1 + \chi^{1-\e} \over 1 + \chi } \biggr] \,, \cr }
\equn\label{Deq8Integral}
$$
where $\chi=t/s$.  The prefactor $\cg$ is defined in \eqn{cgdef}, and
$$
\eqalign{
 {\cal L}(s,t) 
&\equiv 2 (1-2\e) \, (-s) \, (1+\chi) 
   \, (4\pi) \, \I_4^{\oneloop, \, D=6-2\e}[1](s,t) \cr
&= 2i \, \cg \, (-s)^{-\e} \, (1+\chi)    \int_0^1 dx \, {1\over\e} \, 
     { \chi^{-\e} x^{-\e} - (1-x)^{-\e} \over 1-(1+\chi)x } \,, \cr
&= 2i \, \cg \, (-s)^{-\e} \, (1+\chi)    \int_0^1 dx
   \, { 1 \over 1-(1+\chi)x } \biggl[ 
     \ln\left({1-x\over\chi x}\right)
   - {\e\over2} \Bigl( \ln^2(1-x) - \ln^2(\chi x) \Bigr) 
   + \Ord(\e^2) \biggr] \,. \cr }
\equn\label{Ldefexp}
$$
Performing the one-dimensional integral we get,
$$\eqalign{
 {\cal L}(s,t) &= i \, \cg \, (-s)^{-\e} \Biggl\{ \ln^2\chi + \pi^2  \cr
&\hskip 0.5cm
 + \e \biggl[ 2 ( - \Li_3(-\chi) + \zeta(3) 
             + \ln\chi \, \Li_2(-\chi) ) 
             + (\ln^2\chi + \pi^2) \ln(1+\chi) - {2\over3} \ln^3\chi 
             - \pi^2 \ln\chi \biggr] \Biggr\} \cr
&\hskip 3cm
 + \Ord(\e^2). \cr}
\equn\label{Lfinal}
$$
Using \eqns{Deq8Integral}{Lfinal}, the desired one-loop integral is
$$
\eqalign{
\I_4^\oneloop[\mud_p^4](s,t) 
&=  i \, \cg \, (-s)^{-\e} \, (-\e) (1-\e) \, {1\over6} \Biggl\{ 
   {1\over\e}\ 
  - {1\over2} { \chi (\ln^2\chi + \pi^2) \over (1+\chi)^2 } 
  - {\chi\ln\chi\over 1+\chi} + {11\over3}    \cr
&\hskip 1 cm 
 + \e \Biggl[ 
      { \chi\over(1+\chi)^2} \Bigl[ \Li_3(-\chi) - \zeta(3) 
        - \ln\chi \, \Li_2(-\chi) 
        + {1\over3} \ln^3\chi 
        - {1\over2} \ln^2\chi \ln(1+\chi)  \cr
&\hskip 3.5 cm 
        + {\pi^2\over2} \ln\left({\chi\over1+\chi}\right)
        + {1\over2} \left( (2+\chi) \ln^2\chi + \pi^2 \right) \Bigr] \cr
&\hskip 2 cm
    + {11\over3} \biggl(
          - {1\over2} { \chi (\ln^2\chi + \pi^2) \over (1+\chi)^2 }
          - {\chi\ln\chi\over 1+\chi} + {11\over3} \biggr)
          - 4   \Biggr]   \Biggr\} \cr
&\hskip 0.5 cm 
+ \Ord(\e^3), \hskip 2 cm \hbox{region (i).} \cr }
\equn\label{OneLoopBoxResultRegioni}
$$

The three kinematic regions for the four-point amplitude are
\par\noindent
(i) $u>0$ and $s,t < 0$, for which $\chi > 0$;
\par\noindent
(ii) $s>0$ and $t,u < 0$, for which $-1 < \chi <0$; 
\par\noindent
(iii) $t>0$ and $s,u < 0$, for which $\chi < -1$.
\par\noindent
The form~(\ref{OneLoopBoxResultRegioni}) for the one-loop box integral is
appropriate for region (i), where it is manifestly real.  In regions (ii)
and (iii) the integral acquires an imaginary part. 
A form appropriate for region (ii) is obtained by substituting
$(-s)^{-\e} \to s^{-\e} ( 1 + \e \, i \, \pi - \e^2 \, \pi^2/2 )$
and $\ln\chi \to \ln|\chi| + i \, \pi$,
$$
\eqalign{
\I_4^\oneloop[\mud_p^4](s,t) 
&=  i \, \cg \, s^{-\e} \, (-\e) (1-\e) \, {1\over6} \Biggl\{ 
   {1\over\e}\ 
  - {1\over2} { \chi \ln^2|\chi| \over (1+\chi)^2 } 
  - {\chi \ln|\chi| \over 1+\chi} + {11\over3}    \cr
&\hskip 0.5 cm 
 + \e \Biggl[ 
      { \chi \over (1+\chi)^2 } \Bigl[ \Li_3(-\chi) - \zeta(3) 
        - \ln|\chi| \, \Li_2(-\chi) 
        + {1\over3} \ln^3|\chi| 
        - {1\over2} \ln^2|\chi| \ln(1+\chi)  \cr
&\hskip 3.5 cm 
        + {\pi^2\over2} \ln|\chi|
        + {1\over2} (2+\chi) \ln^2|\chi| \Bigr] \cr
&\hskip 2 cm
    - { \pi^2 \over 2 (1+\chi) } 
    + {11\over3} \biggl(
          - {1\over2} { \chi \ln^2|\chi| \over (1+\chi)^2 }
          - {\chi\ln|\chi|\over 1+\chi} + {11\over3} \biggr)
          - 4   \Biggr]  \cr
&\hskip 0.5 cm
+ i \, \pi \Biggl( 
     - {\chi\ln|\chi| \over (1+\chi)^2} + {1 \over 1+\chi} \cr
&\hskip 1.2 cm 
 + \e \Biggl[ { \chi \over (1+\chi)^2 } \Bigl[ \Li_2(1+\chi)
              + {1\over2} \ln^2|\chi| + \ln|\chi| \Bigr]
   - {11\over3} \biggl( { \chi \ln|\chi| \over (1+\chi)^2 }
                      - {1 \over 1+\chi} \biggr) \Biggr] \Biggr)
\Biggr\} \cr
&\hskip 0.5 cm
 + \Ord(\e^3), \hskip 2 cm \hbox{region (ii).} \cr }
\equn\label{OneLoopBoxResultRegionii}
$$
A form appropriate for region (iii) can be obtained simply by applying
the transformation $s \lr t$ ($\chi \lr 1/\chi$) to
\eqn{OneLoopBoxResultRegionii}, since that transformation maps 
region (ii) into region (iii).

We now return to the non-vanishing planar double box integral.  
Its expression in terms of $\I_4^\oneloop[\mud_p^4]$ is,
$$
\I_4^\P[\mud_{p+q}^2 \mud_q^2](s,t)\ =\  
-i \cg \, {1\over\e^2} \, (-s)^{-1-\e} \, \I_4^\oneloop[\mud_p^4](s,t) 
\ +\ { F^\P_{p+q,q} \over (4\pi)^4 \, (-s) }\ +\ \Ord(\e),
\equn\label{PlanarMinusSing}
$$
where the finite remainder is
$$
\eqalign{
F^\P_{p+q,q}\ &=\  {1\over18} {\chi\over(1+\chi)^2} \Bigl[
      - \ln\chi \, (\ln^2\chi + \pi^2) 
      + \Bigl(\chi-{1\over\chi}\Bigr) \, \pi^2 \Bigr] \,, 
\hskip 3 cm \hbox{region (i),}  \cr
&=\ {1\over18} {\chi\over(1+\chi)^2} \Bigl[
      - \ln|\chi| \, (\ln^2|\chi| - 2 \pi^2) 
      + \Bigl(\chi-{1\over\chi}\Bigr) \, \pi^2 
- 3 i \pi \ln^2|\chi| \Bigr]  \,, 
\hskip 0.25 cm \hbox{regions (ii) and (iii).}  \cr }
\equn\label{FPlanar}
$$
In rewriting \eqn{PlanarInts}, we used 
$$
\ctg = \cg^2 \Bigl[ 1 + \Ord(\e^3) \Bigr]. 
\equn\label{ctgcgRelation}
$$
We remark that the apparent power-law singularities as $\chi \to -1$ in 
eqs.~(\ref{OneLoopBoxResultRegionii}) and (\ref{FPlanar}) are all spurious, 
cancelling among the various terms.

The analytic structure of the non-planar integrals is more intricate than
that of the planar ones.  This is because, for Mandelstam variables
$s,t,u$ satisfying $s+t+u=0$, there is no Euclidean region where the
integrals are purely real.  (It is possible to find a Euclidean region by
relaxing the condition $s+t+u=0$, but this introduces another
dimensionless variable and leads to more cumbersome 
expressions~\cite{Tausk}.)  Also in contrast to the planar
case, the non-planar integrals do not have a uniform $i\ve$ prescription
in terms of the dimensionless ratio $\chi$.  The correct $i\ve$ 
prescription can be determined by following through the $i\ve$s appearing 
in the Feynman propagators.

We express the divergent non-planar integral in terms of the
one-loop box integral, as in \eqn{PlanarMinusSing},
$$
\I_4^\NP[\mud_p^2 \mud_q^2](s,t)\ =\  
-i \cg \, {1\over\e^2} \, (-s)^{-1-\e} \, \I_4^\oneloop[\mud_p^4](u,t) 
\ +\ { F^\NP_{p,q} \over (4\pi)^4 \, (-s) }\ +\ \Ord(\e).
\equn\label{NonPlanarMinusSing}
$$
Here the finite remainder is
$$
\eqalign{
F^\NP_{p,q} &= {1\over6} \Biggl\{ 
 - 2 \chi(1+\chi) \biggl[ 
         \Li_3\Bigl({\chi\over1+\chi}\Bigr) - \zeta(3) 
       - \ln\Bigl({\chi\over1+\chi}\Bigr)
          \Bigl( \Li_2\Bigl({\chi\over1+\chi}\Bigr) + {\pi^2\over2} \Bigr)
       - {1\over6} \ln^3\Bigl({\chi\over1+\chi}\Bigr) \biggr] \cr
& \hskip 0.5 cm
 + 3 \chi(1+\chi) \ln(1+\chi) \ln\chi
 - {1\over2} (1+\chi)^2 \biggl( -{1\over\chi} + 3 \biggr) \ln^2(1+\chi)
 - {1\over2} \chi^2 \biggl( {1\over1+\chi} + 3 \biggr) \ln^2\chi \cr
& \hskip 0.5 cm
 + \pi^2 \biggl( \chi - {1\over2} {1\over 1+\chi} 
              + {5\over6} \biggr)
 + (1+\chi) \ln(1+\chi) - \chi \ln\chi  \cr
& \hskip 0.1 cm
+ i \pi \biggl( 
    2 \chi(1+\chi) \biggl[ \Li_2\Bigl({\chi\over1+\chi}\Bigr) 
                            - {\pi^2\over6} - {3\over2} \ln\chi  \biggr]
  + (1+\chi) \biggl[ 
         (1+\chi) \Bigl( -{1\over\chi} + 3 \Bigr) \ln(1+\chi) - 1 \biggr]
  \biggr) \Biggr\} \,, \cr
& \hskip 4 cm
\hbox{region (i),} \cr}
\equn\label{FNonPlanarRegioni}
$$
$$
\eqalign{
F^\NP_{p,q} &= {1\over6} \Biggl\{ 
 - 2 \chi(1+\chi) \biggl[ 
         \Li_3\Bigl({\chi\over1+\chi}\Bigr) - \zeta(3) 
       - \ln\Bigl\vert{\chi\over1+\chi}\Bigr\vert
            \Li_2\Bigl({\chi\over1+\chi}\Bigr) 
       - {1\over6} \ln^3\Bigl\vert{\chi\over1+\chi}\Bigr\vert \biggr] \cr
& \hskip 0.5 cm
 + 3 \chi(1+\chi) \Bigl( \ln(1+\chi) \ln|\chi| - {\pi^2\over2} \Bigr)
 - {1\over2} (1+\chi)^2 \biggl( -{1\over\chi} + 3 \biggr) \ln^2(1+\chi) \cr
& \hskip 0.5 cm
 - {1\over2} \chi^2 \biggl( {1\over1+\chi} + 3 \biggr) \ln^2|\chi|
 + {\pi^2\over3} + (1+\chi) \ln(1+\chi) - \chi \ln|\chi|  \cr
& \hskip 0.1 cm
+ i \pi \biggl( 
   \chi(1+\chi) \biggl[ \ln^2\Bigl\vert{\chi\over1+\chi}\Bigr\vert 
                        + \pi^2 \biggr]
  - (1+\chi) \Bigl( -{1\over\chi} + 2 \Bigr) \ln(1+\chi) 
  + \chi \Bigl( {1 \over 1+\chi} + 2 \Bigr) \ln|\chi| 
  + 1 \biggr) \Biggr\} \,, \cr
& \hskip 4 cm
\hbox{region (ii),} \cr}
\equn\label{FNonPlanarRegionii}
$$
and the form for region (iii) may be obtained from that for region (i) 
through the transformation $t \lr u$ ($\chi \lr -1-\chi$).
The region (ii) form for $F^\NP_{p,q}$ is symmetric under 
$\chi \lr -1-\chi$, although the identities~\cite{Lewin}
$$
\eqalign{
 \Li_3(-x) - \Li_3(-1/x) &= -{\pi^2\over6} \ln x - {1\over6} \ln^3x \,,
\hskip 1 cm x > 0\,,\cr
 \Li_2(-x) + \Li_2(-1/x) &= -{\pi^2\over6} - {1\over2} \ln^2x \,, 
\hskip 1.7 cm x > 0\,, \cr}
\equn\label{TrilogDilogId}
$$
are needed to demonstrate this.

Finally, the integral 
$\I_4^\NP[\mud_{p+q}^2 \mud_q^2] = \I_4^\NP[\mud_{p+q}^2 \mud_p^2]$
is finite, but non-vanishing, as $\e\to0$.  It is given by
$$
\eqalign{
\I_4^\NP[\mud_{p+q}^2 \mud_q^2] & =  {1\over (4 \pi)^4(-s)} 
        {1\over 6}  \Biggl\{ 
 {\chi \over (1+\chi)^2} \biggl[  \Li_3(-\chi) - \zeta(3) 
          - \ln\chi \Bigl( \Li_2(-\chi) - {\pi^2\over6} \Bigr)
          - {3\over4} \chi \bigl( \ln^2\chi - \pi^2 \bigr) \biggr] \cr
& \hskip 0.5 cm
 - {1+\chi \over \chi^2} \biggl[  \Li_3\Bigl({1\over1+\chi}\Bigr) - \zeta(3) 
          + \ln(1+\chi) \Bigl( \Li_2\Bigl({1\over1+\chi}\Bigr) 
                             + {\pi^2\over6} \Bigr)
          + {3\over4} (1+\chi) \ln^2(1+\chi) \cr
& \hskip 2.5 cm
          + {1\over3} \ln^3(1+\chi) \biggr]
 + \Bigl( {1\over \chi(1+\chi)} + {3\over2} \Bigr) 
                    \ln(1+\chi) \ln\chi \cr
& \hskip 0.5 cm
 + \pi^2 \biggl( {1 \over 6\chi} + {4 \over 3(1+\chi)}
               + {3\over2} {\chi\over(1+\chi)^2} - {3\over4} \biggr)
 + {\ln(1+\chi)\over 2\chi} - {\ln\chi\over 2(1+\chi)}   \cr
& \hskip 0.1 cm 
 + i \pi \biggl( 
 - { 1+\chi \over \chi^2 } \biggl[ 
     \Li_2\Bigl( { \chi \over 1+\chi } \Bigr)
     - \ln(1+\chi) \ln\chi  
     + {1\over2} \ln^2(1+\chi)  
     - {3\over2} (1+\chi) \ln(1+\chi) \biggr]
\cr & \hskip 1.5 cm
    - {1\over2} {\chi\over(1+\chi)^2} 
      \bigl( \ln^2\chi + \pi^2 \bigr)
    - \biggl( {1 \over \chi(1+\chi)} + {3\over2} \biggr) \ln\chi
    - {1\over2\,\chi}\ \biggr)  \Biggr\} \,, \cr
& \hskip 4 cm
\hbox{region (i),} \cr}
\equn\label{NonsingNonplanarRegioni}
$$
$$
\eqalign{
\I_4^\NP[\mud_{p+q}^2 \mud_q^2] & =  {1\over (4 \pi)^4(-s)} 
        {1\over 6}  \Biggl\{ 
 {\chi \over (1+\chi)^2} \biggl[  \Li_3(-\chi) - \zeta(3) 
          - \ln|\chi| \Bigl( \Li_2(-\chi) + {5\pi^2\over6} \Bigr)
          - {3\over4} \chi \ln^2|\chi| \biggr] \cr
& \hskip 0.5 cm
 + {1\over 2} \Bigl( {1\over \chi(1+\chi)} + {3\over2} \Bigr) 
                    \ln(1+\chi) \ln|\chi|
 - \pi^2 \biggl( {5 \over 12 \, \chi(1+\chi)} + {3\over8} \biggr)
 - {\ln|\chi| \over 2(1+\chi)}   \cr
& \hskip 0.1 cm 
 + i \pi \biggl( 
 - { \chi \over (1+\chi)^2 } \biggl[ 
     \Li_2\Bigl( { 1+\chi \over \chi } \Bigr)
     - \ln(1+\chi) \ln|\chi| - {3\over2} \ln|\chi| \biggr]
\cr & \hskip 1.5 cm
    - {1 \over 1+\chi} \biggl[ \ln(1+\chi) 
           - {1\over2} \ln|\chi| + {1\over2} \biggr] \biggr) 
\ +\ (\chi \lr -1-\chi)\ \Biggr\} \,, \cr
& \hskip 4 cm
\hbox{region (ii).} \cr}
\equn\label{NonsingNonplanarRegionii}
$$
Again the form for region (iii) may be obtained from that for region (i) 
through the transformation $t \lr u$ ($\chi \lr -1-\chi$).


\end{document}